\newif\ifarxiv
\arxivtrue 

\ifarxiv
	\documentclass[12pt]{article}
	\usepackage[total={6.5in, 9in}]{geometry}
	\usepackage{natbib}
	\usepackage{tgtermes}
	\usepackage[T1]{fontenc}
	\usepackage[urw-garamond]{mathdesign}
\else
	\documentclass[review]{elsarticle}
	\usepackage{lineno,hyperref}
	\modulolinenumbers[5]
	\journal{Transportation Reseacrh Part C}
	\usepackage{amsfonts}
\fi

\usepackage{graphicx,float}
\usepackage{amsmath}

\usepackage{color,url}

\DeclareMathOperator*{\mini}{minimize}
\DeclareMathOperator*{\argmin}{argmin}

\graphicspath{{./}{./fig/}}
\newcommand{\pred}[2]{x_{#1}^{#2}}
\newcommand{\myx}{x}

\newcommand{\y}{y}

\begin{document}
\ifarxiv
\author{
	Nicholas G. Polson\\
	\textit{Booth School of Business}\\
	\textit{University of Chicago}\footnote{Polson is Professor of Econometrics and Statistics
		at the Chicago Booth School of Business. email: ngp@chicagobooth.edu. Sokolov is an assistant professor at George Mason University, email: vsokolov@gmu.edu}\\
	Vadim O. Sokolov\\
	\textit{Systems Engineering and Operations Research}\\
	\textit{George Mason University}\\
}
\else
\begin{frontmatter}
		\author{Nicholas G. Polson}
		\address{Booth School of Business, University of Chicago, Chicago, Illinois, 60637, U.S.A. }
		
		\author{Vadim O. Sokolov\corref{mycorrespondingauthor}}
		\address{Systems Engineering and Operations Research, George Mason University, Fairfax, Virginia, 22030, U.S.A. }
		\cortext[mycorrespondingauthor]{Corresponding author}
		\ead{vsokolov@gmu.edu}
\fi
\title{Deep Learning for Short-Term Traffic Flow Prediction}

\ifarxiv
\date{First Draft: December 2015\\
	This Draft: February 2017
}
\maketitle
\else
\end{frontmatter}
\fi

\begin{abstract}   
	We develop a deep learning model to predict traffic flows. The main contribution is development of an architecture that combines a linear model that is fitted using $\ell_1$ regularization and a sequence of $\tanh$ layers. The challenge of predicting traffic flows are the sharp nonlinearities due to transitions between  free flow,  breakdown, recovery and  congestion. We show that deep learning architectures can capture these nonlinear spatio-temporal effects.  The first layer identifies spatio-temporal relations among predictors and other layers model nonlinear relations.  We illustrate our methodology on road sensor data from Interstate I-55 and predict traffic flows during two special events; a Chicago Bears football game and an extreme snowstorm event. Both cases have sharp traffic flow regime  changes, occurring very suddenly, and we show how  deep learning provides precise short term traffic flow predictions. 
\end{abstract}
	
%
%
%
%
%
%
%

\section{Introduction}
\subsection{Traffic Flow Prediction}
Real-time  spatio-temporal measurements of traffic flow speed are available from  in-ground loop detectors or GPS probes. Commercial traffic data providers, such as Bing maps~\cite{clearflow},  rely on traffic flow data, and machine learning to predict speeds for each road segment. Real-time  (15-40 minute) forecasting gives travelers the ability to choose better routes and authorities the ability to manage the transportation system. Deep learning is a form of machine learning which provides good short-term forecasts of traffic flows by exploiting the dependency in the high dimensional set of explanatory variables, we capture the sharp discontinuities in traffic flow that arise in large-scale networks. We provide a variable selection methodology based on sparse models and dropout.


The goal of our paper is to model the nonlinear spatio-temporal effects in recurrent and non-recurrent traffic  congestion patterns. These arise due to conditions at construction zones, weather, special events, and traffic incidents. Quantifying travel time uncertainty requires real-time forecasts. Traffic managers use model-based forecasts to regulate ramp metering, apply speed harmonization, and regulate road pricing as a congestion mitigation strategy; whereas,  the general public adjusts travel decisions on departure times and travel route choices, among other things. 

Deep learning forecasts congestion propagation given a bottleneck location, and can provide an accurate forty minute forecasts for days with recurrent and non-recurrent traffic conditions.  Deep learning can also incorporate other  data sources, such as weather forecasts, and police reports to produce more accurate forecasts.  We illustrate our methodology on traffic flows during two special events;  a Chicago Bears football game and an extreme snow storm event.

To perform variable selection, we develop a hierarchical sparse vector auto-regressive technique \cite{dellaportas2012joint, nicholson2014structured} as the first deep layer.  Predictor selection then proceeds in a dropout~\cite{hinton2006reducing}. Deep learning models  the sharp discontinuities in traffic flow are modeled as a superposition of univariate non-linear activation functions with affine arguments. Our procedure is scalable and estimation follows traditional optimization techniques, such as stochastic gradient descent.

The rest of our paper is outlined as follows. Section~\ref{sec:existing-work} discusses connections with existing work. Section~\ref{sec:int-dl} reviews fundamentals of deep learning.  Section \ref{sec:dl} develops deep learning predictors for forecasting traffic flows. Section~\ref{sec:chi} discusses fundamental characteristics of traffic flow data and illustrates our methodology with the study of traffic flow on Chicago's I-55. Finally, Section \ref{sec:discussion} concludes with directions for future research.

\subsection{Connections with Existing Work}\label{sec:existing-work}
Short-term traffic flow prediction has a long history in the transportation literature. Deep learning is a form of machine learning that can be viewed as a nested hierarchical model which includes traditional  neural networks. \cite{karlaftis2011statistical} provides an overview of traditional neural network approaches and \cite{kamarianakis2012real} shows that model training is computationally expensive with frequent updating being prohibitive. On the other hand, deep learning with dropout can find a sparse model which can be frequently updated in real time.   There are several analytical approaches to traffic flows modeling \cite{anacleto2013multivariate,bla12, chiou2013,polson2014bayesian,polson2014bayesiana, work2010traffic}. These approaches can  perform very well on filtering  and state estimation. The caveat is that they are hard to implement on large scale networks. Bayesian approaches have been  shown to be efficient for handling large scale transportation network state estimation problems \cite{tebaldi1998bayesian}. \cite{westgate2013travel} discusses ambulance  travel time reliability  using noisy GPS for both path travel time and individual road segment travel time distributions. \cite{anacleto2013multivariate} provides a dynamic Bayesian network to model external intervention techniques to accommodate situations with suddenly changing traffic variables. 

Statistical and machine learning methods for traffic forecasting are compared in ~\cite{smith1997traffic}. \cite{sun_bayesian_2006} provides a Bayes network algorithm, where the conditional probability of a traffic state on a given road, given states on topological neighbors on a road network is calculated. The resulting joint probability distribution is a mixture of Gaussians. Bayes networks for estimating travel times were suggested by \cite{horvitz2012prediction} which eventually became a commercial product that led to the start of Inrix, a traffic data company.  \cite{wu_travel-time_2004} provides a  machine-learning method support vector machine (SVM)~\cite{polson2011data} to forecast travel times and \cite{quek_pop-traffic:_2006} proposes a fuzzy neural-network approach to address nonlinearities in traffic data. \cite{rice_simple_2004} argues that there is a linear relation between future travel times and currently estimated conditions with a time-varying coefficients regression model to predict travel times.

Integrated  auto-regressive moving average (ARIMA) and exponential smoothing (ES) for traffic forecasting are studied in \cite{tan_aggregation_2009} and \cite{van1996combining}. A Kohonen self-organizing map is proposed as an initial classifier. \cite{van_lint_online_2008} addresses real-time  parameter learning and improves the quality of forecasts using an extended Kalman filter. \cite{ban2011real} proposes a method for estimating queue lengths at controlled intersections, based on the travel time data measured by GPS probes. The method relies on detecting discontinuities and changes of slopes in travel time data.  \cite{ramezani2015queue} combines the traffic flow shockwave analysis with data mining techniques.  \cite{oswald2000traffic} argues that non-parametric methods produce better forecasts than parametric models due to their ability to better capture spatial-temporal relations and non-linear effects.  \cite{vlahogianni2014short} provides an extensive recent review of literature on short-term traffic predictions. 

There are several issues not addressed in the current literature  \cite{vlahogianni2014short}. First, predictions at a network level using data-driven approaches. There are two situations when a data-driven approach might be preferable to methodologies based on traffic flow equations. Estimating boundary conditions is a challenging task, which even in systems that rely on loop detectors as traffic sensors are typically not installed on ramps.  Missing data  problems are usually addressed using data imputation~\cite{muralidharan2009imputation} or  weak formulations of boundary conditions~\cite{s2006weak}. Our results show that a data-driven approach can efficiently forecast flows without boundary measurements from ramps.  Another challenge with physics-based approaches comes from their limited ability to model urban arterials. For example,  \cite{qiao2001intelligent} shows  analytical approaches fail to provide good forecasts. Another challenge is to identify spatio-temporal relations in flow patterns,  \cite{vlahogianni2014short} for further discussion.  Data-driven approaches provide a flexible alternative to physical laws of traffic flows.  

The challenge is to perform model selection  and residual diagnostics  \cite{vlahogianni2014short}. Model selection can be  tackled by regularizing the loss function and using cross-validation to select the optimal penalty weight. To address this issue, when we specify our deep learning model we construct an architecture as follows. First we use is a regularized vector autoregressive model to perform predictor selection. Then, our deep learning model addresses the issue of non-linear and non-stationary relations between variables (speed measurements) using a series of activation functions.


Breiman \cite{breiman2003statistical} describes the trade-off between machine learning and  traditional statistical methods. Mchine learning has been widely applied \cite{brian1996pattern} and shown to be particularly successful in traffic pattern recognition. For example, shallow neural networks for traffic applications \cite{chen2001use}, use a memory efficient dynamic neural network based on resource allocating network (RAN)  with a single hidden layer with Gaussian radial basis function  activation unit. \cite{zheng2006short} develops several one-hidden layer networks to produce  fifteen-minute forecasts. Two types of networks, one with a $\tanh$ activation function and one with a Gaussian radial basis function were developed. Several forecasts were combined using a Bayes factors that calculates an odds ratio for each of the models dynamically. \cite{van2005accurate} proposes a state-space neural network and a multiple hypothesis approach that relies on using several neural network models at the same time  \cite{van2009bayesian}. Day of the week and  time of day as inputs to a neural network was proposed in \cite{ccetiner2010neural}. Our work is closely related to \cite{lv2015traffic}, which  demonstrates that deep learning can be effective for traffic forecasts.  A stacked auto-encoder  was used to learn the spatial-temporal patterns in the traffic data with training  performed by a greedy layer-wise fashion. \cite{ma2015long} proposed a recurrent architecture, a Long Short-Term Memory Neural Network (LSTM), for travel speed prediction. Our approach builds on this by showing an additional advantage of deeper hidden layers together with sparse autoregressive techniques for variable selection.

\subsection{Deep Learning}\label{sec:int-dl}
Deep learning learns a high dimensional function via a sequence of semi-affine non-linear transformations. The deep architecture is organized as a graph. The nodes of the graph are units, connected by links to propagate activation, calculated at the origin, to the destination units. Each link has a weight that determines the relative strength and sign of the connection and each unit  applies an activation function to all of the weighted sum of incoming activations. The activation function is given, such as a hard threshold, a sigmoid function or a $\tanh$. A particular class of deep learning models uses a directed acyclic graph structure is called a feed-forward neural network.  There is vast literature on this topic; one of the earlier works include \cite{bishop1995neural} \cite{haykin2004comprehensive}.

Deep learning allows for efficient  modeling of nonlinear functions, see the original problem of  Poincare and Hilbert. The  advantage of  deep hidden layers is for a high dimensional input variable, $\myx = (x_1, \ldots, x_p)$ is that  the activation functions are univariate, which implicitly requires the specification of the number of hidden units $N_l$ for each layer $l$. 

The Kolmogorov-Arnold representation theorem \cite{kolmogorov1956representation} provides the theoretical motivation for deep learning. The theorem states that any continuous function of $n$ variables, defined by $F(x)$, can be represented as
\[
F(x) = \sum_{j=1}^{2n+1}g_j\left(\sum_{i=1}^{n}h_{ij}(x_i)\right),
\]
where $g_j$ and $h_{ij}$ are continuous functions, and $h_{ij}$ is a universal basis, that does not depend on $F$. This remarkable representation result implies that any continuous function can be represented using operations of summation and function composition. For a neural network, it means that any function of $n$ variables can be represented as a neural network with one hidden layer and $2n+1$ activation units. The difference between theorem and neural network representations is that functions $h_{ij}$ are not necessarily   affine. Much research has focused on how to find such a basis. In their original work, Kolmogorov and Arnold develop  functions in a constructive fashion. \cite{diaconis1984nonlinear} characterizes projection pursuit functions for a specific types of  input functions.

%

A  deep learning predictor, denoted by $\hat{y}(x)$, takes an input vector $x = (x_1,\ldots, x_p)$ and outputs $y$ via  different layers of abstraction that employ hierarchical predictors by composing $L$ non-linear semi-affine transformations. Specifically, a deep learning architecture is as follows. Let $f_1,\ldots f_n$ be given univariate activation link functions, e.g. sigmoid ($1/(1+e^{-x})$, $\cosh(x)$, $\tanh(x)$), Heaviside gate functions ($I(x > 0)$), or rectified linear units ($\max\{x, 0\}$) or indicator functions ($I(x \in R)$) for trees. The composite map is defined by
$$
\hat{y}(x):= F(x)  = \left ( f_{w_n,b_n} \circ \ldots \circ f_{w_1,b_1} \right ) ( \myx),
$$
where $f_{w,b}$ is a semi-activation rule defined by
\begin{equation}\label{eq:deep-function}
f_{w_l,b_l} (x) =  f \left ( \sum_{j=1}^{N_l} w_{lj} \myx_j + b_l \right ) = f ( w_l^T \myx_l + b_l )\quad (l=1,\ldots, n).
\end{equation}
Here $N_l$ denotes the number of units at layer $l$. The weights $w_l \in R^{N_l \times N_{l-1}}$ and offset $b\in R$ needs to be learned from training data.

Data dimension reduction of a high dimensional map $F$ is performed via the composition of univariate semi-affine functions. Let $z^l$ denote the $l$-th layer hidden features, with $x = z^0$. The final output is the response $y$,  can be numeric or categorical. The explicit structure of a deep prediction rule is than
\begin{align*}
z^1 = &f(w_0^T \myx + b_0)\\
z^2 = &f(w_1^T z^1 + b_1)\\
&\cdots\\
z^q = &f(w_{n-1}^T z^{n-1} + b_{n-1})\\
y(x) =& w_n^T z^n + b_n.
\end{align*}

In many cases there is an underlying probabilistic models, denoted by $p(y \mid \hat{y}(x))$. This leads to a training problem given by optimization problem
\[
\min_{w,b} \quad\dfrac{1}{T}\sum_{i=1}^{T}-\log p(y_i\mid \hat{y}_{w,b}(x_i)),
\]
where $p(y_i\mid \hat{y}(x))$ is the probability density function given by specification
$
y_i = F(x_i) + \epsilon_i.
$
For example, if $\epsilon_i$ is normal, we will be training $\hat{w},~\hat{b}$ via an $\ell_2$-norm, $\min_{w,b} ||y - F_{w,b}(x)||^2 = \sum_{i=1}^{T}(y_i - F_{w,b}(x_i))^2$. One of the key advantages of deep learning is the the derivative information $\nabla_{w,b} l(y,\hat{y}_{w,b}(x))$ is available in closed form via the chain rule. Typically, a regularization penalty, defined by $\lambda\phi(w,b)$ is added, to introduce the bias-variance decomposition to provide good out-of-sample predictive performance. An optimal regularization parameter, $\lambda$, can be chosen using out-of-sample cross-validation techniques. One of the advantages of $\ell_1$ penalized least squares formulation is that it leads to a convex, though non-smooth, optimization problem. Efficient algorithms \cite{kim2007} exist to solve those problems, even for high dimensional cases.

There is a strong connection with nonlinear multivariate non-parametric models, which we now explore. In a traditional statistical framework, the non-parametric approach seeks to approximate the unknown  map $F$ using a family of functions defined by the following expression
\begin{equation}\label{eqn:approx}
F(x)  = \sum_{k=n}^{N} w_k f_k(x).
\end{equation}
Functions $f_k$ are called basis functions and play the similar role of a functional space basis, i.e. they are chosen to give a good approximation to the unknown map $F$. In some cases $\{f_k\}_{k=1}^{N}$ actually do form a basis of a space, e.g., Fourier ($f_k(x)  = \cos(kx)$) and wavelet bases. Multi-variate basis functions are usually constructed using functions of a single variable. Four examples are radial functions, ridge functions, kernel functions and indicator functions. 
\begin{equation}
f_k(x) = 
\begin{cases}
\kappa\left(||X - \gamma_k||_2\right) \mbox{ (radial function)}\\
\kappa(w^TX + w_0) \mbox{ (ridge function)}\\
\kappa\left(\frac{X-\gamma_k}{h}\right) \mbox{ (kernel estimator)}\\
I(X \in C_k) \mbox{ (tree indicator function)} 
\end{cases}
\label{eqn:basis}
\end{equation}

Here $\kappa$ is typically chosen to be a bell-shaped function (e.g., $1/e^{x^2}$ or $1/\cosh(x)$). The ridge function, a composition of inner-product and non-linear univariate functions, is arguably one of the  simplest non-linear multi-variate function. Two of the most popular types of neural networks are constructed as a composition of radial or ridge functions. Popular non-parametric  \textit{tree-based models}~\cite{breiman1984classification} can be represented as (\ref{eqn:approx}) by choosing $f_k$ given by equation \ref{eqn:basis}. In tree-based regression, weights  $\alpha_k = \bar{Y}_k$ are the averages of $(Y_i \mid X_i \in C_k)$ and $C_k$ is a box set in $R^p$ with zero or more extreme directions (open sides). 



Another set of basis functions are \textit{Fourier series}, used primarily for time series analysis, where $f_k(x) = \cos (x)$. A \textit{spline} approximation can also be derived by using polinomial functions with finite support as a basis. 

Ridge-based models, can efficiently represent high-dimensional data sets with a small number of parameters. We can think of deep features (outputs of hidden layers) as projections of the input data into a lower dimensional space. Deep learners can deal with the curse of dimensionality because ridge functions determine  directions in $(z^{k-1}, z^{k})$ input space, where the variance is very high. Those directions are chosen as global ones and represent the most significant  patterns in the data. This approach resembles  the other well-studied techniques such as projection pursuit~\cite{friedman1974projection}  and principal component analysis.

\section{Deep Learning for Traffic Flow Prediction} \label{sec:dl}
Let $\pred{t+h}{t}$  be the forecast of traffic flow speeds at time $t+h$, given measurements up to time $t$. Our deep learning traffic architecture looks like 
\[
y(x): = \pred{t+40}{t} = \left(\begin{array}{c}
x_{1,t+40}\\ \vdots\\
x_{n,t+40}
\end{array} \right).
\]

To model traffic flow data $x^t = (x_{t-k},\ldots,x_t)$ we use predictors $x$ given by
\[
x^t = \mathrm{vec} \left(\begin{array}{ccc}
x_{1,t-40}&...&x_{1,t}\\ \vdots & \vdots& \vdots\\
x_{n,t-40}&... &x_{n,t}
\end{array} \right).
\]
Here $n$ is the number of locations on the network (loop detectors) and $x_{i,t}$ is the cross-section traffic flow speed at location $i$ at time $t$.  We use, \verb|vec| to dennote the vectorization transformation, which converts the matrix into a column vector. In our application examined later in Section~\ref{sec:traffic}, we use twenty-one road segments (i.e.,  $n=21$) which span thirteen miles of a major corridor connecting Chicago's southwest suburbs to the central business district. The chosen length is consistent with several current transportation  corridor management deployments~\cite{transnet}.

Our layers are constructed as follows, $x = z^{0}$, then  $z^{l+1}$, $l=0, \ldots ,L$ is a time series ``filter" given by
\begin{align*}
z_i^{l+1} =& f\left(\sum_{i=1}^{N_l}\left(w_i^{l+1}z_i^{l} + b_i^{l+1}\right)\right) \quad (i=1,\ldots,N_{l+1}).
\end{align*}




As we show later in out empirical studies, the problem of non-linearities in the data is efficiently addressed by the deep learners. 

\begin{figure}
	\begin{center}
		\includegraphics[width=1\linewidth]{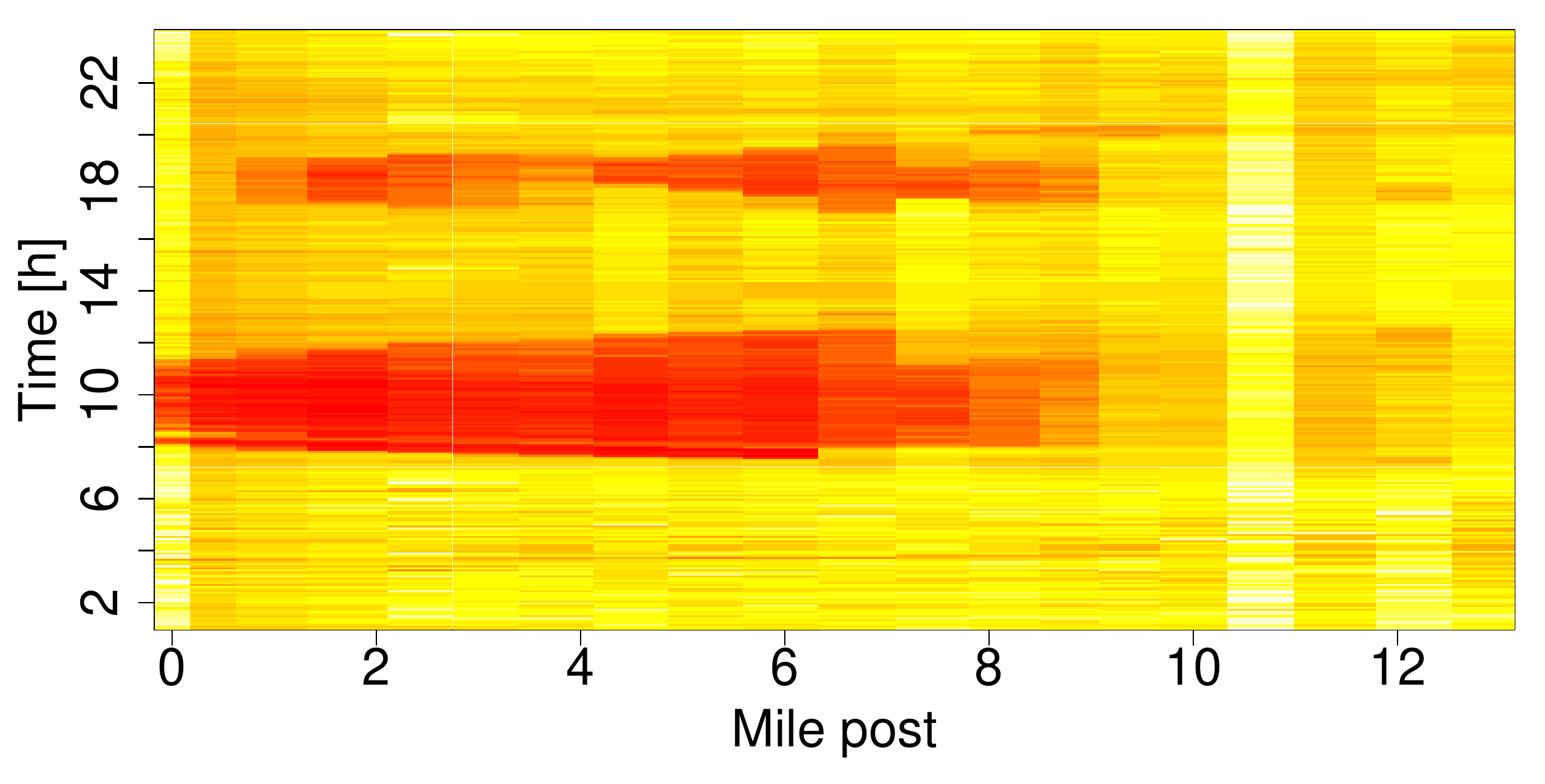}
	\end{center}
	\vspace*{-15pt}
	\caption{A time-space diagram that shows traffic flow speed on a 13-mile stretch of I-55. Cross-section speed measured on from 18 February 2009 (Wednesday). \textcolor{red}{Red} means slow speed and light \textcolor{yellow}{yellow} corresponds to free flow speed. The direction of the flow is from 0 to 13.}
	\label{fig:heat}
\end{figure}

The problem of finding the spatio-temporal relations in the data is the predictor selection problem. Figure  \ref{fig:heat} shows a time-space diagram of traffic flows on a 13-mile stretch of highway I-55 in Chicago. You can see a clear spatio-temporal pattern in traffic congestion propagation in both downstream and upstream directions. 

A predictor selection problem requires an algorithm to find a sparse models. Those rely on adding a penalty term to the loss function. A recent review by \cite{nicholson2014structured} considers several  prominent scalar regularization terms to identify sparse vector auto-regressive models. 

Our approach will be to develop a hierarchical linear vector autoregressive model to identify the spatio-temporal relations in the data. We consider the problem of finding sparse matrix, $A$, in the following model
\[
\pred{t+40}{t} = A x^t  + \epsilon_t, \quad \epsilon_t\sim N(0,V);
\]
where $A$ is a matrix of size $n\times nk$, and $k$ is the number of previous measurements used to develop a forecast.  In our example in Section~\ref{sec:traffic}, we have $n=21$; however, in large scale networks, there are tens of thousands locations with measurements available.

The predictors selected as a result of finding the sparse linear model at the first layer are then used to build a deep learning model. To find an optimal network (structure and weights) we used the stochastic gradient descent (SGD) method implemented in the package \verb|H2O|. Similar methods are available in \verb|Python|'s \verb|Theano| \cite{Bastien-Theano-2012} or \verb|TensorFlow| \cite{abadi2016tensorflow} framework.  We use random search to find meta parameters of the deep learning model. To illustrate our methodology, we generated $N=10^5$ Monte Carlo samples from the following space:
\begin{align*}
f& \in \{\tanh(x), ~\max(x,0)\}\\
n  &\in \{1,\ldots,60\}\\
N_l   &\in \{1,\ldots,200\}^{n}\\
\lambda         &\in [10^{-4}, ~10^{-2}]\\
\pred{t+h}{t} &=  \left ( f_n \circ \ldots \circ f_1 \right )( \myx^t), \quad f_l =   f ( w_l^T \myx_l + b_l ).
\end{align*}
We used out-of-sample model performance as a criteria for selecting our  final deep learning architecture.

\subsection{Training}
At a fundamental level, we use a training set $(y_i, x_i)_{i=1}^N$ of input-output pairs, to train our deep learning model by minimizing the difference  between training target $y_i$ and our predictor $\hat{y}(x_i)$. To do this, we require a loss function, $l(\y,\hat{\y})$ at the level of the output signal that measures our goodness-of-fit. When we have a traditional probabilistic model $p(y \mid \hat{y})$ that generates the output $y$ given the predictor $\hat{y}$, than we have the natural loss
function $l(y, \hat{y} ) = - \log p( y~|~\hat{y} )$. For deep learning problems, a typical $\ell_2$ norm squared used as a loss function. It is common to add a regularization penalty $ \phi(w,b)$ that will avoid over-fitting and stabilize our predictive rule. To summarize, given an activation function, the statistical problem is to optimally find the weights and biases $w = (w_0,\ldots,w_n)$,  $b = (b_0,\ldots,b_n)$ that minimize the loss function with $\ell_2$ separable regularization term given by
\begin{align*}
(\hat{w},~\hat{b}) \in& \argmin_{w,b} ~ \Vert \y - \hat{\y}_{w,b} (\myx) \Vert_2^2 + \lambda \phi(w,b)\\
\phi(w,b) = & ||w||^2 + ||b||^2\\
\hat{\y}_{w,b} (\myx) = &\left ( f_n \circ \ldots \circ f_1 \right )( \myx^t), \quad f_l(x) =    f \left ( \sum_{j=1}^{n_l} w_{lj} \myx_j + b_l \right ),
\end{align*}
here $w_l \in \mathbb{R}^{n_l}$, $b_l \in \mathbb{R}$,  $ $ and  $ \lambda$ gages the overall level of regularization. Choices of penalties for $ \phi(w,b) $ include the ridge penalty $\ell_2$ or the lasso $\ell_1$ penalty to induce sparsity in the weights. A typical method to solve the optimization problem is  stochastic gradient descent with mini-batches. The caveat of this approach include poor treatment of multi-modality and  possibly slow convergence. From our experiments with the traffic data, we found that using sparse linear model estimation to identify spatial-temporal relations between variables yields  better results, as compared to using dropout or regularization terms for neural network loss function~\cite{pachitariu2013regularization,helmbold2015inductive}. Both penalized fitting and dropout \cite{srivastava2014dropout} are techniques for preventing overfitting. They control the bias-variance trade-off and improve out-of-sample performance of a model. A regularized model is less likely to overfit and will have a smaller size. Dropout  considers all possible combinations of parameters by randomly dropping a unit out. Instead of considering different networks with different units being dropped out, a single network is trained with some probability of each unit being dropped out. Then during testing, the weights are scaled down according to drop-out probability to ensure that expected output is the same as actual output at the test time. Dropout is a heuristic method that is little understood but was shown to be very successful in practice. There is a deep connection though between penalized fitting for generalized linear models and dropout technique. \cite{wager2013} showed that dropout is a first order equivalent to $\ell_2$ penalization after scaling the input data by the Fisher information matrix. For traffic flow prediction, there are predictors that are irrelevant, such as changes in traffic flow in a far away location that happened five minutes ago do not carry any information. Thus, by zeroing-out those predictors, the $\ell_1$ penalization leads to the  sparsity pattern that better capture the spatio-temporal relations. A similar observation has been made in \cite{kamarianakis_real}.

In order to find an optimal structure of the neural network (number of hidden layers $L$,  number of activation units in each layer $N_l$ and activation functions $f$) as well as hyper-parameters, such as $\ell_1$ regularization weight, we used a random search. Though this technique can be inefficient for large scale problems, for the sake of exploring potential structures of the networks that deliver good results and can be scaled, this is an appropriate technique for small dimensions. Stochastic gradient descent used for training a deep learning model scales linearly with the data size. Thus the hyper-parameter search time is linear with respect to model size and data size. On a modern processor it takes about two minutes to train a deep learning network on 25,000 observations of 120 variables. To perform a hyper-parameter and model structure search we fit the model $10^5$ times. Thus the total wall-time (time that elapses from start to end) was 138 days. An alternative to random search for learning the network structure for traffic forecasts was proposed in \cite{vlahogianni2005optimized} and relies on the genetic optimization algorithm.

\subsection{Trend Filtering}
\begin{figure}
	\begin{center}
		\begin{tabular}{ccc}
			\includegraphics[width=0.33\linewidth]{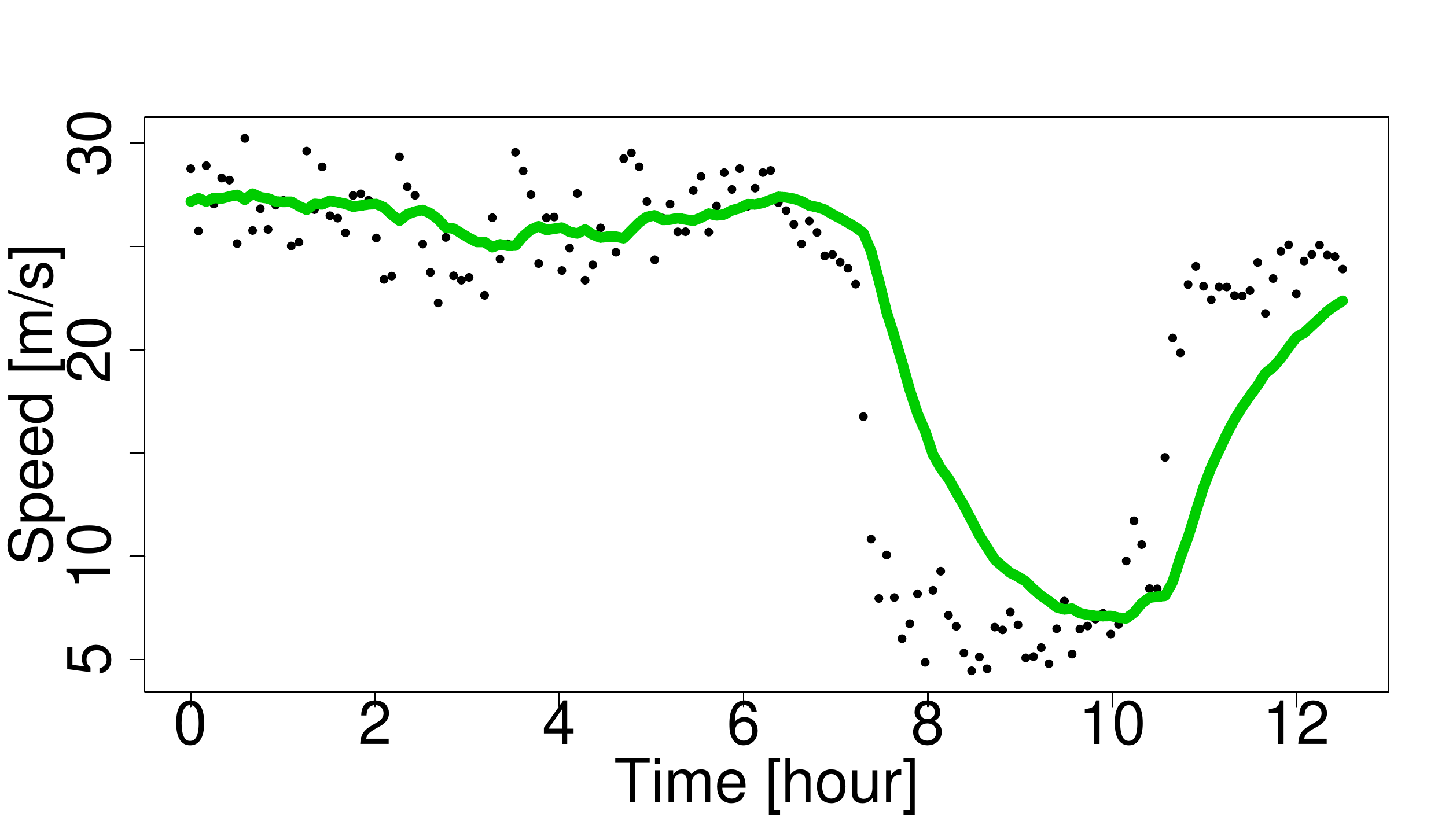} &
			\includegraphics[width=0.33\linewidth]{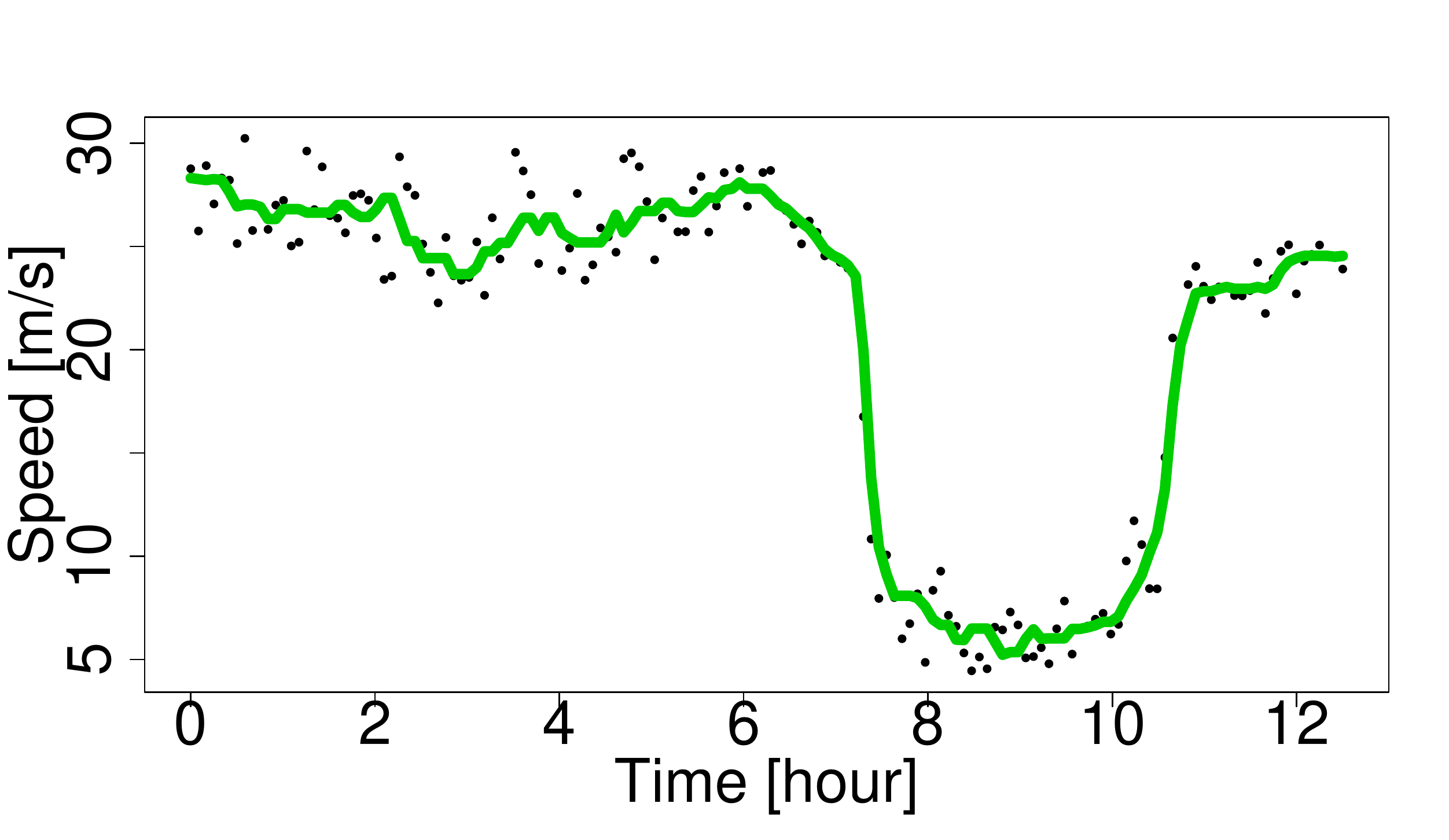} &
			\includegraphics[width=0.33\linewidth]{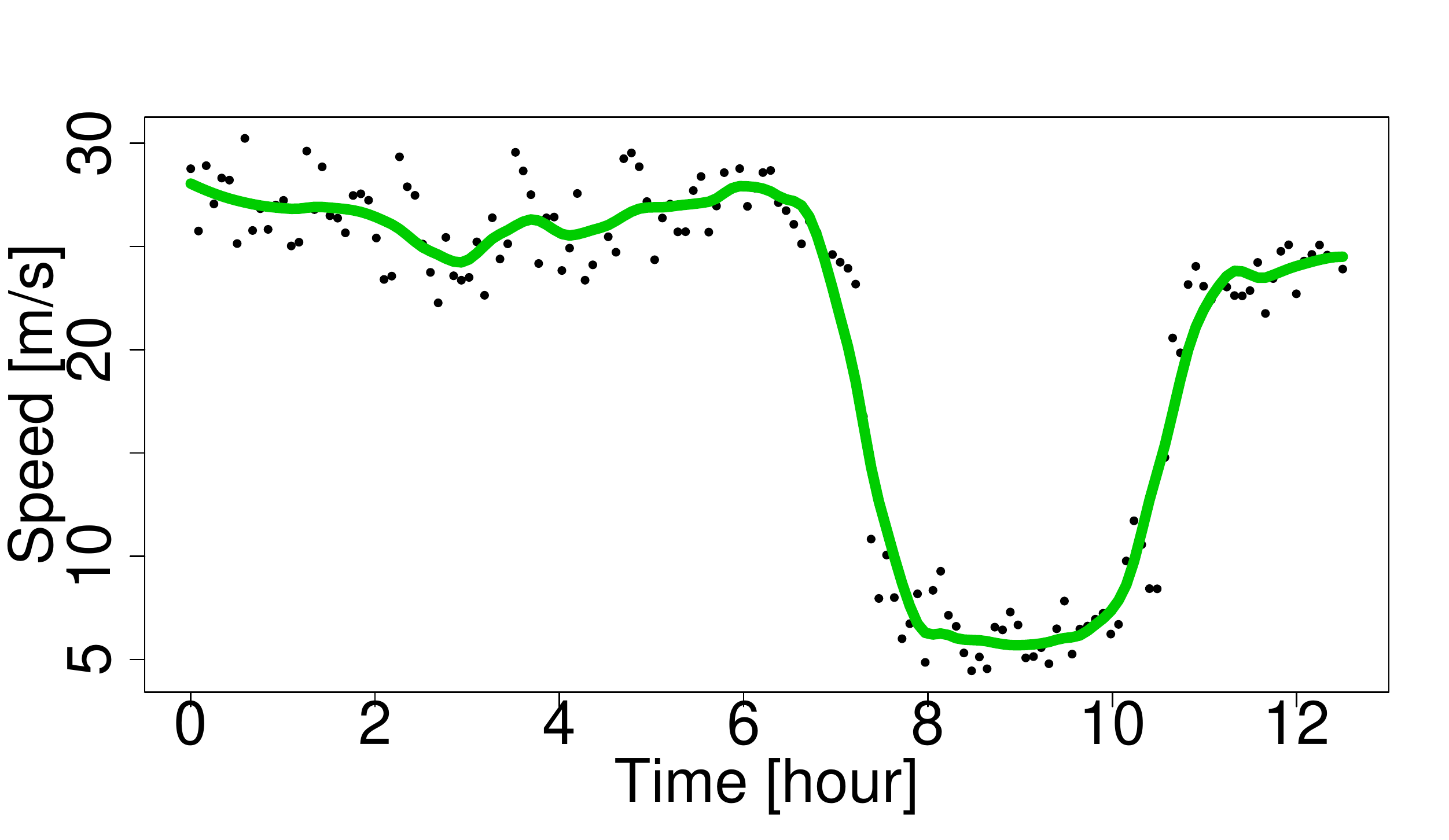} \\
			(a) exponential smoothing & (b) median filter & (c) loess filter
		\end{tabular}
	\end{center}
	\caption{Results of three classical filtering techniques applied to traffic data (cross-section speed) from one of the work days.}
	\label{fig:ets}
\end{figure}

One goal of traffic flow modeling is to filter noisy data from physical sensors and then to develop model-based predictors. Iterative exponential smoothing is a popular technique that is computationally efficient. It smoothers  oscillations that occur on arterial roads with traffic signal controls, when measured speed is ``flipping'' between two values, depending whether the  probe vehicle stopped at the light or not. Figure \ref{fig:ets}(a), however, shows that it does not work well for quickly switching regimes observed in highway traffic. Another approach is median filtering, which unlike exponential smoothing captures quick changes in regimes as shown on Figure~\ref{fig:ets}(b). However, it will not perform well on an arterial road controlled by traffic signals, since it will be oscillating back and forth between two values.  A third approach is to use a piecewise polynomial fit to filter the data. Figure \ref{fig:ets}(c) shows that this method does perform well, as the slopes  be underestimated. Median filtering seems to be the most effective in our context of traffic flows.

An alternative approach to filter the data is to assume that we observe data from the statistical model $v_i = f(x_i) + e_i$ , where $f(x)$ is piecewise constant. The fused lasso \cite{tibshirani2011solution} and $\ell_1$ trend filtering  \cite{kim2009ell_1, polson2014mixtures} involves estimating $f(x) = (f(x_1),\ldots,f(x_n))$ at the input points by solving the optimization problem
\[
\mini_{} \quad ||y - f(x) ||_2^2 + \lambda ||Df(x)||_1.
\]  
In fused lasso $D = D^{(1)}$ is the matrix encoding first differences in $f(x)$. In $\ell_1$ trend filtering $D = D^{(2)}$ is the matrix encoding second differences in $f(x)$ 
\[
D^{(1)} = \left(
\begin{array}{cccccc}
1 & -1 & 0 & 0 & \cdots & 0 \\ 
0 & 1 & -1 & 0 & \cdots & 0 \\ 
\vdots &  &  &  & \ddots & \vdots \\ 
0 & \cdots &  & 0 & 1 & -1
\end{array} 
\right),\quad
D^{(2)} = \left(
\begin{array}{cccccc}
1 & -2 & 1 & 0 & \cdots & 0 \\ 
0 & 1 & -2 & 1 & \cdots & 0 \\ 
\vdots &  &  &  & \ddots & \vdots \\ 
0 & \cdots &  & 1 & -2 & 1
\end{array} 
\right).
\]

Applying $D^{(1)}$ to a vector is equivalent to calculating first order differences of the vector. This filter also called 1-dimensional total variation denoising \cite{rudin1992nonlinear}, and hence first order $\ell_1$ trend filtering estimate $f(x)$ is piecewise constant across the input points $x_1,\ldots,x_n$. Higher orders difference operators $D^{(k)}$, $k>1$, correspond to an assumption that the data generating process is modeled by a piece-wise polynomial of order $k$ function.

The non-negative parameter, $\lambda$, controls the trade-off between smoothness of the signal and closeness to the original signal. The objective function is strictly convex and thus can be efficiently solved to find a unique minimizer $x^{\mathrm{lt}}$. The main reason to use the trend filtering is that it produces a piece-wise linear function in $t$. There are integer times, $1 = t_1 < t_2,..., < t_p = n$ for which 
\[
x^{\mathrm{lt}} = \alpha_k + \beta_k t,~~t_k\le t \le t_{k+1},~~k=1,\ldots,p-1
\]
Piece-wise linearity is guaranteed by using the $\ell_1$-norm penalty term. It guarantees sparseness of $Df(x)$ (the second-order difference of
the estimated trend), i.e.  it will have many zero elements, which means that the estimated trend is piecewise linear. The points $t_2,\ldots,t_{p-1}$ are called kink points. The kink points correspond to change in slope and intercept of the estimated trend and can be interpreted as points at which regime of data generating process changes. This function well aligns with the traffic data from an in-ground sensor. The regimes in data correspond to free flow, degradation, congestion and recovery. Empirically, the assumption that data is piecewise linear is well justified. Residuals of the filtered data $x^{\mathrm{lt}} - y$ are typically low and show no patterns.  

A trend filter is similar to a spline fitting method with one important difference. When we fit a spline (piece-wise continuous polynomial function) to the data, we need to provide knots (kinks) as inputs. Trend filtering has the advantage that the kinks and parameters of each line are found jointly. 

%

\begin{figure}
	\begin{center}
		\includegraphics[width=0.6\linewidth]{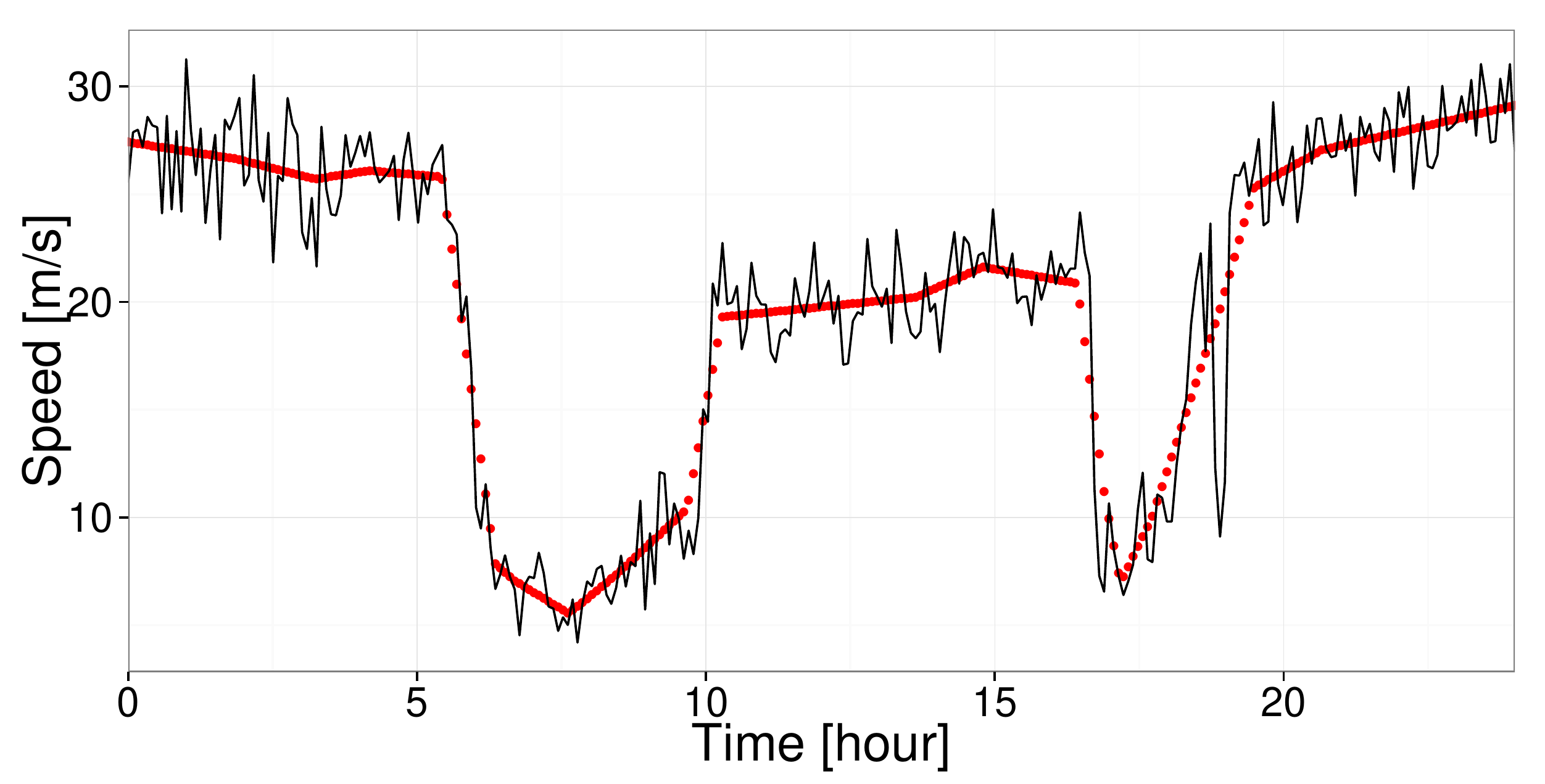}
	\end{center}
	\caption{$\ell_1$ trend filtering based on quadratic loss and penalty that enforces a piecewise line fit applied to one day of measured cross-section speed.}
	\label{fig:tf}
\end{figure}

Figure~\ref{fig:tf} shows results of applying $\ell_1$ trend filter to a data measured from a loop-detector on I-55. 
A computationally efficient algorithms for trend filtering with differentiation operator of any order $D^{(k)}$ was recently proposed by \cite{ramdas2015fast}.

\section{Chicago Traffic Flow During Special Events}\label{sec:chi}

To illustrate our methodology, we use data from twenty-one loop-detectors installed on a northbound section of Interstate I-55 . Those loop-detectors span 13 miles of the highway.  Traffic flow data is available from the Illinois Department of Transportation, (see Lake Michigan Interstate Gateway Alliance  \url{http://www.travelmidwest.com/}, formally the Gary-Chicago-Milwaukee Corridor, or GCM). The  data is measured by  loop-detector sensors installed on interstate highways. Loop-detector is a  simple presence sensors that measure when a vehicle is present and generate an on/off signal. There are over 900 loop-detector sensors that cover the Chicago metropolitan area.  Since 2008, Argonne National Laboratory has been archiving traffic flow data every five minutes from the grid of sensors. Data contains averaged \textit{speed}, \textit{flow}, and \textit{occupancy}.  Occupancy is defined as percent of time a point on the road is occupied by a vehicle, and flow is the number of off-on switches. Illinois uses a single loop detector setting, and speed is estimated based on the assumption of an average vehicle length.

A distinct  characteristic of  traffic flow data is an abrupt change of the mean level. Also we see a lot of variations on the speed measurements. Though, on Figure \ref{fig:ets}, it might seem that during the congested period (6am -- 9am) the speed variation is small; in reality the signal to noise ratio during congested regime is lower compared to a free flow regime. One approach to treat the noisy data is a probabilistic one. In \cite{polson2014bayesian} the authors develop a hierarchical Bayesian model for tracking traffic flows and estimate uncertainty about state variables at any given point in time.  However, when information is presented to a user, it has to be presented as a single number, i.e. travel time from my origin to my destination is 20 minutes. A straightforward way to move from a distribution over a traffic state variable (i.e., traffic flow speed) to a single number is to calculate an expected value or a maximum a posteriori.

\subsection{Traffic Flow on Chicago's Interstate I-55}\label{sec:traffic}

One of the key attributes of congestion propagation on a traffic network is the spatial and temporal dependency between bottlenecks. For example, if we consider a stretch of highway and assume a bottleneck, than it is expected that the end of the queue will move from the bottleneck downstream. Sometimes, both the head and tail of the bottleneck move downstream together. Such discontinuities in traffic flow, called shock waves are  well studied and can be modeled using a simple flow conservation principles. However, a similar phenomena can be observed not only between downstream and upstream locations on a highway. A similar relationship can be established between locations on  city streets and highways \cite{horvitz2012prediction}. 

\begin{figure}
	\centering
	\begin{tabular}{cc}
		\includegraphics[width=0.5\linewidth]{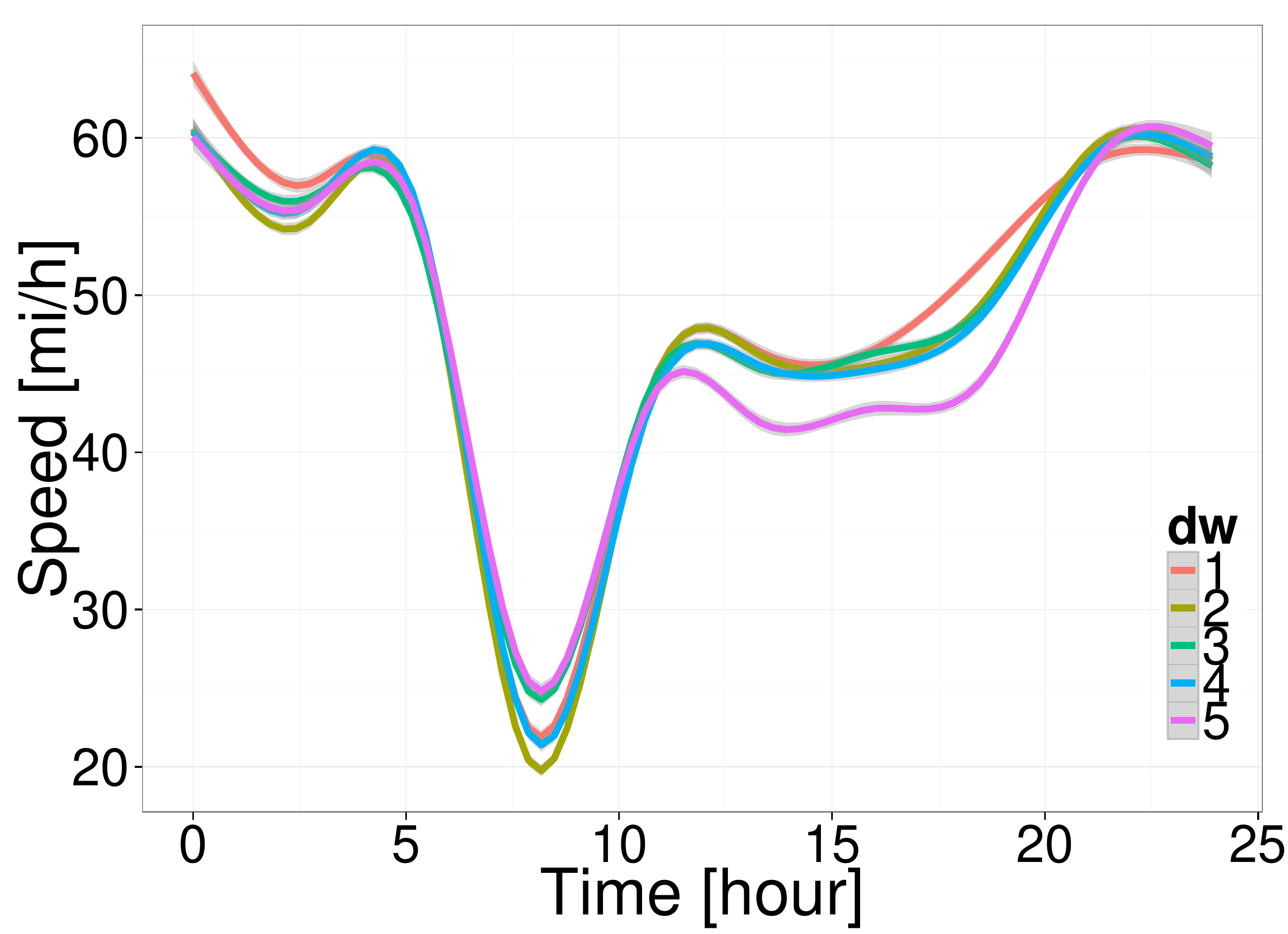} &
		\includegraphics[width=0.5\linewidth]{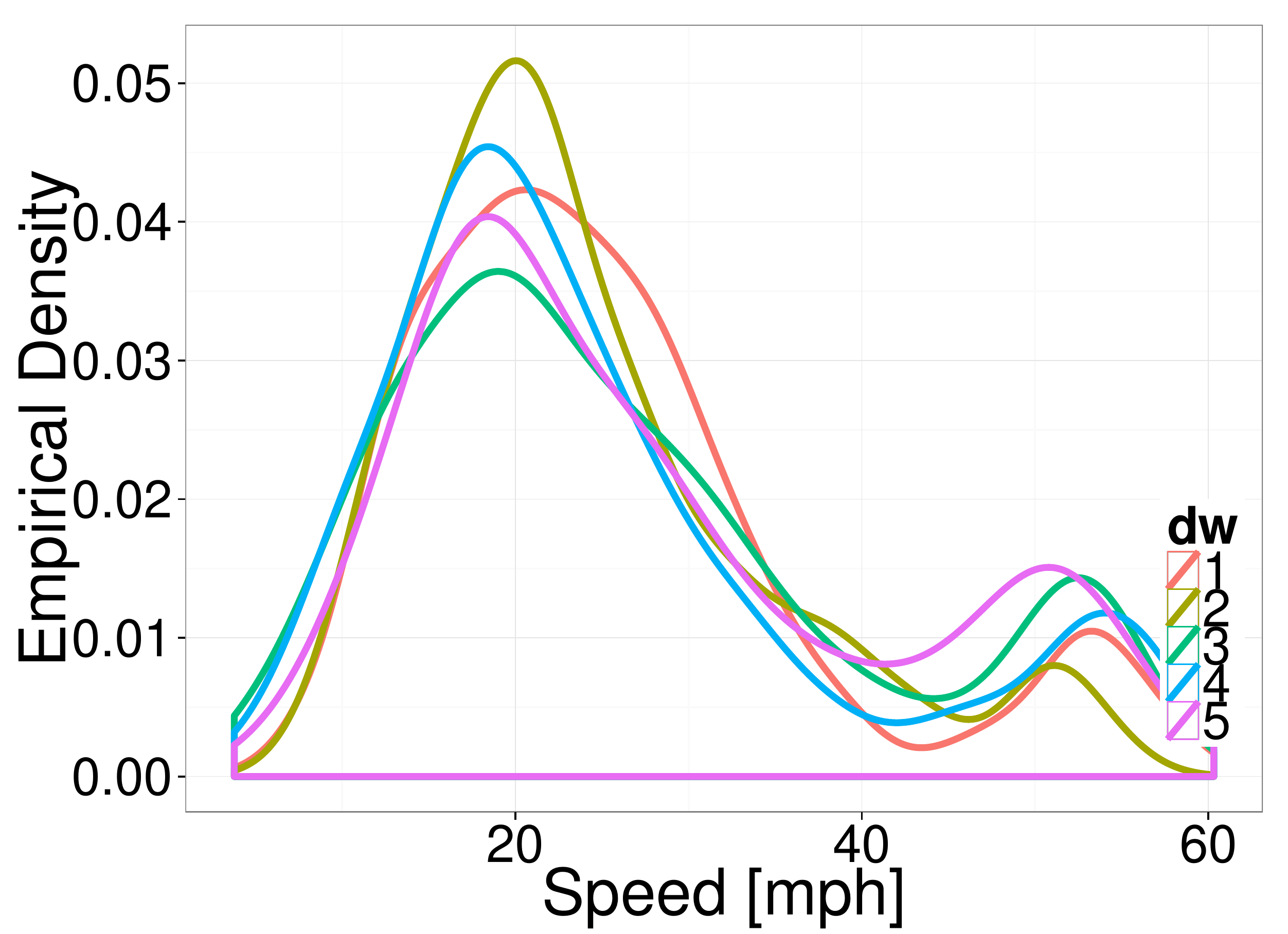}\\
		(a) Average speed on work days & (b) Empirical density for speed, on work days
	\end{tabular}
	\caption{Traffic patterns on different days of the week. Left panel (a) shows  average speed on work days. Right panel (b) shows empirical density for speed, for five work days of the week. Calculated based on the data collected between 7 and 8am.}
	\label{fig:five-day}
\end{figure}

Another important aspect of traffic congestion is that it can be ``decomposed'' into recurrent and non-recurrent factors. For example, a typical commute time from a western suburb to Chicago's city center on Mondays is 45 minutes. However, occasionally the travel time is 10 minutes shorter or longer.  Figure \ref{fig:five-day} shows summarized data collected from the sensor located eight miles from the Chicago downtown on I-55 northbound, which is part of a route used by many morning commuters to travel from southwest suburbs to the city.  Figure~\ref{fig:five-day}(a) shows average speed on the selected road segment for each of the five work days; we can see that there is  little variation, on average, from one week day to another with travelers most likely to experience delays between 5 and 10am. However, if we look at the empirical probability distribution of travel speeds between 7 and 8 am on the same road segment on Figure \ref{fig:five-day}(b), we see that distribution is bi-modal. In most cases, the speed is around 20 miles per hour, which corresponds to heavy congestion. The free flow speed on this road segment is around 70 miles per hour. Furthermore, the distribution has a heavy left tail. Thus, on many days the traffic is considerably worse, compared to an ``average'' day. 

\begin{figure}[H]
	\centering
	\begin{tabular}{cc}
		\includegraphics[width=0.5\linewidth]{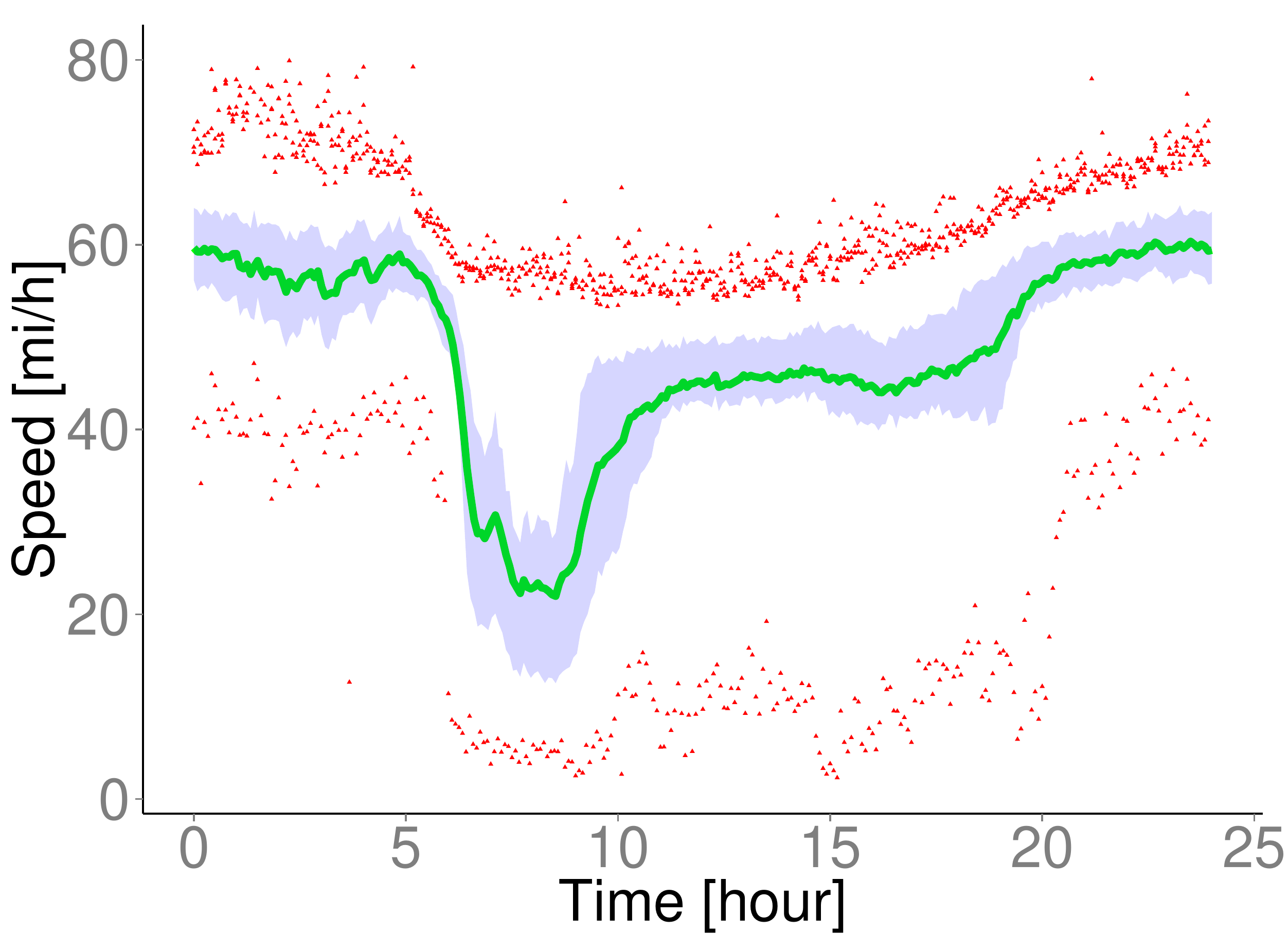} &
		\includegraphics[width=0.5\linewidth]{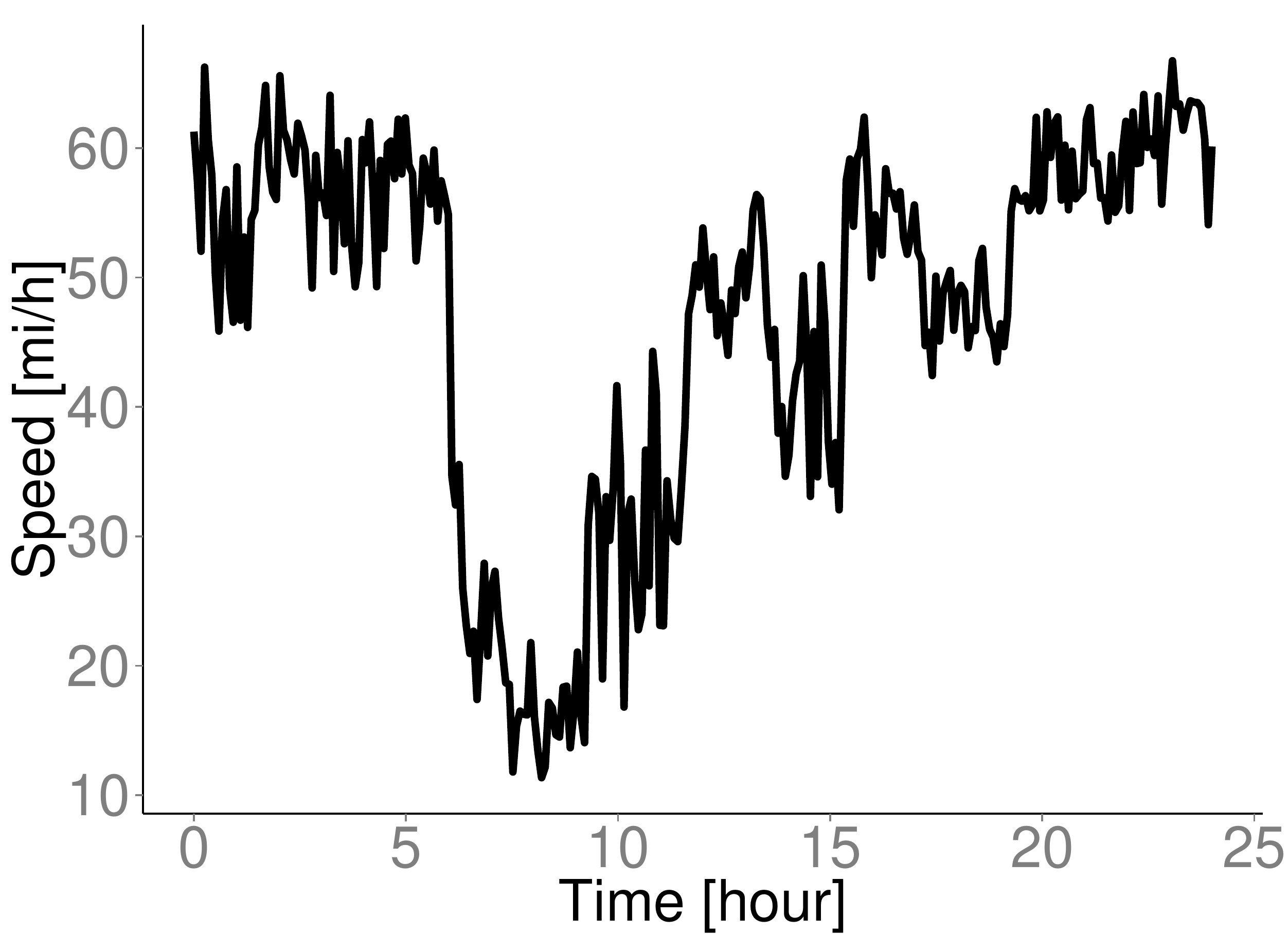}\\
		(a) Speed measured on Thursdays & (b) Example of one day speed profile
	\end{tabular}
	\caption{Recurrent speed profile. Both plots show the speed profile for a segment of interstate highway I-55. Left panel (a) shows the green line, which is the average cross-section speed for each of five minute intervals with 60\% confidence interval. The red points are measurements that lie outside of 98\% confidence interval. Right panel (b)  shows an example of one day speed profile from May 14, 2009 (Thursday). }
	\label{fig:wed}
\end{figure}

Figure~\ref{fig:wed}(a) shows measurements from all non-holiday Wednesdays in 2009.  The solid line and band, represent the average speed and 60\% confidence interval correspondingly. Each dot is an individual speed measurement that lies outside of 98\% confidence interval. Measurements are taken every five minutes, on every Wednesday of 2009; thus, we have roughly 52 measurements for each of the five-minute intervals. 
We see that in many cases traffic patterns are very similar from one day to another. However, there are many days when we see surprises, both good and bad.  A good surprise might happen, e.g., when schools are closed due to extremely cold weather. A bad surprise might happen due to non-recurrent traffic conditions, such as an accident or inclement weather.

Figure~\ref{fig:wed}(b)  illustrates a typical day's traffic flow pattern on Chicago's I-55 highway. This traffic pattern is recurrent, we can see a breakdown in flow speed during the morning peak period, followed by speed recovery. The free flow regimes are usually of little interest to traffic managers.

\begin{figure}
	\begin{tabular}{cc}
		\includegraphics[width=0.5\textwidth]{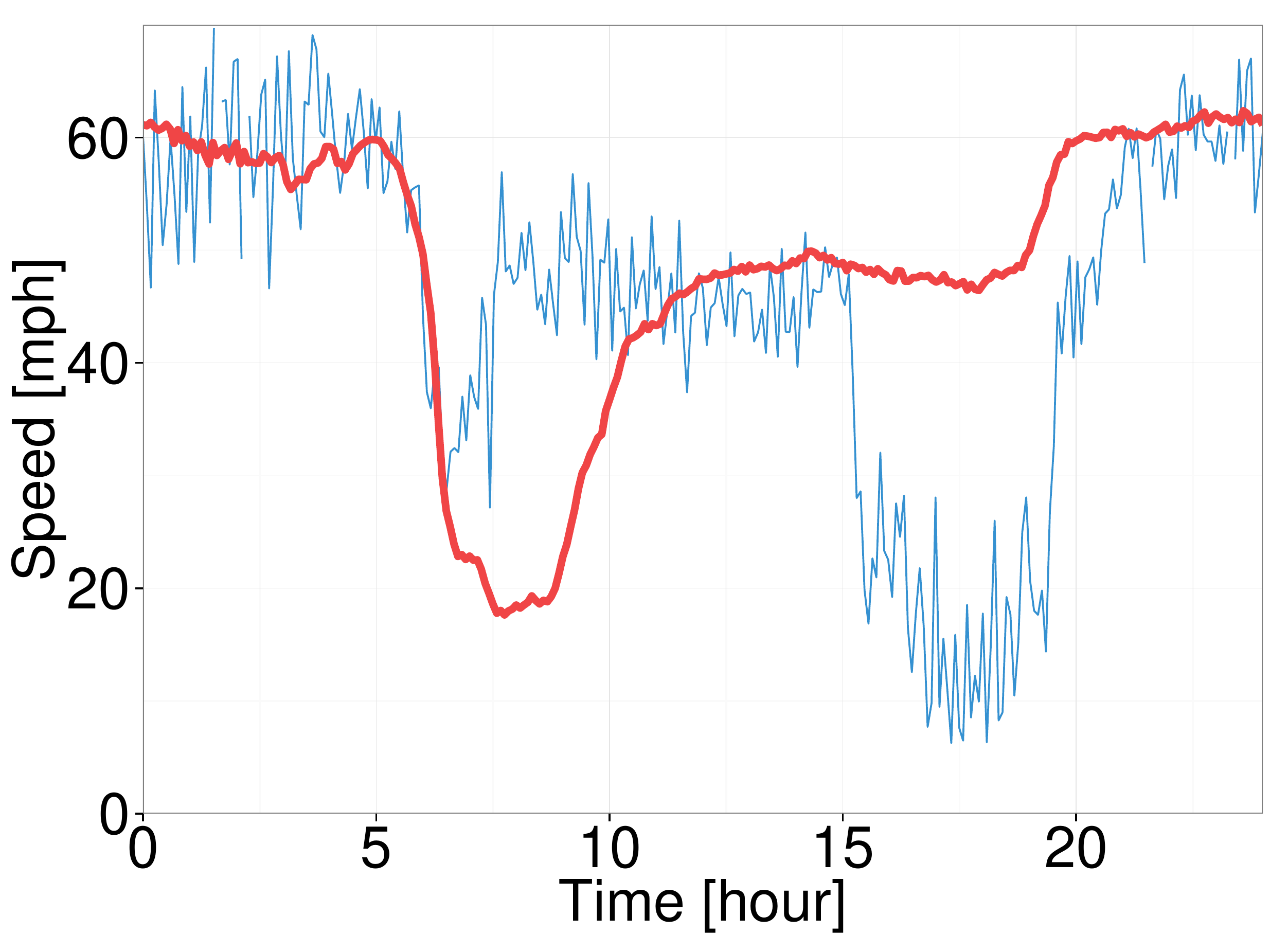} & 
		\includegraphics[width=0.5\linewidth]{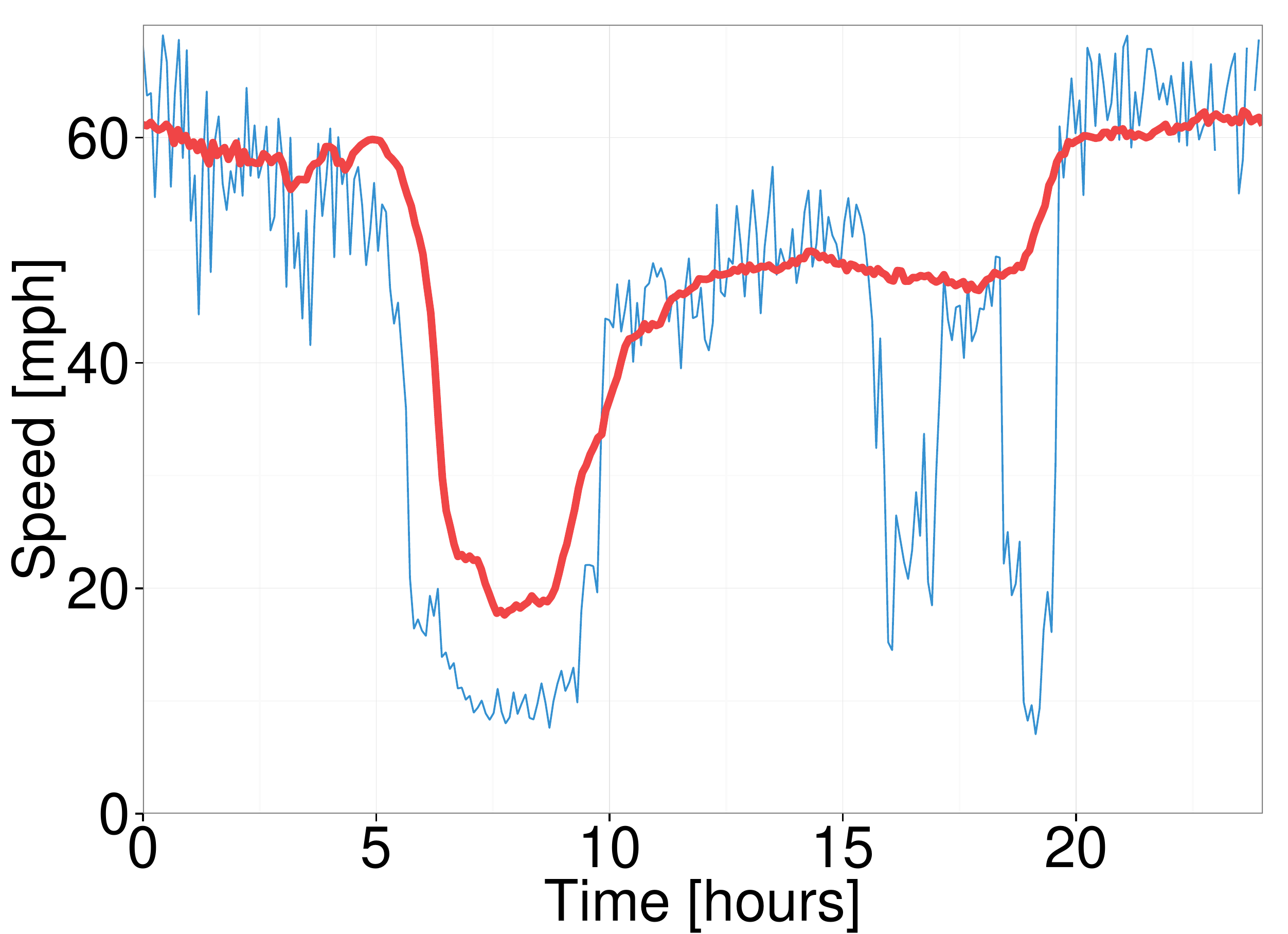}\\
		(a) Chicago Bears football game & (b) Snow weather
	\end{tabular} 
	\caption{Impact of non-recurrent events on traffic flows. Left panel (a) shows traffic flow on a day when New York Giants played at Chicago Bears on Thursday October 10, 2013.  Right panel (b) shoes impact of light snow on traffic flow on I-55 near Chicago on December 11, 2013. On both panels average traffic speed is red line and speed on event day is blue line.}
	\label{fig:special-events}
\end{figure}

Figure \ref{fig:special-events} shows the impact of non-recurrent events. In this case, the traffic speed can significantly deviate  from historical averages due to the increased number of vehicles on the road or due to poor road surface conditions. 
Our goal is to build a statistical model to capture the sudden regime changes from free flow to congestion and then the decline in speed to the recovery regime for both recurrent traffic conditions and non-recurrent ones.

\begin{figure}[H]
	\begin{tabular}{cc}
		\includegraphics[width=0.45\linewidth]{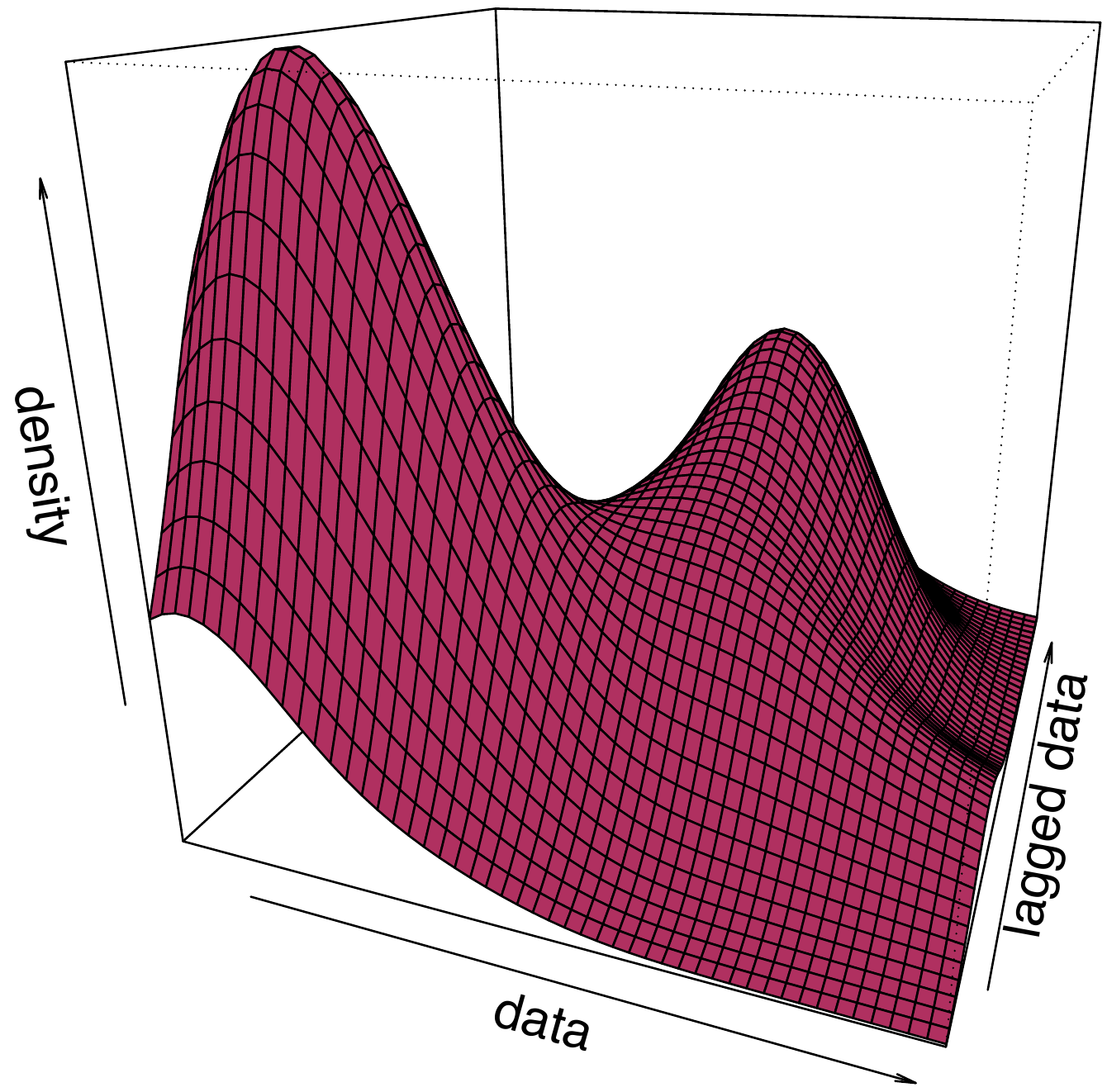} &
		\includegraphics[width=0.45\linewidth]{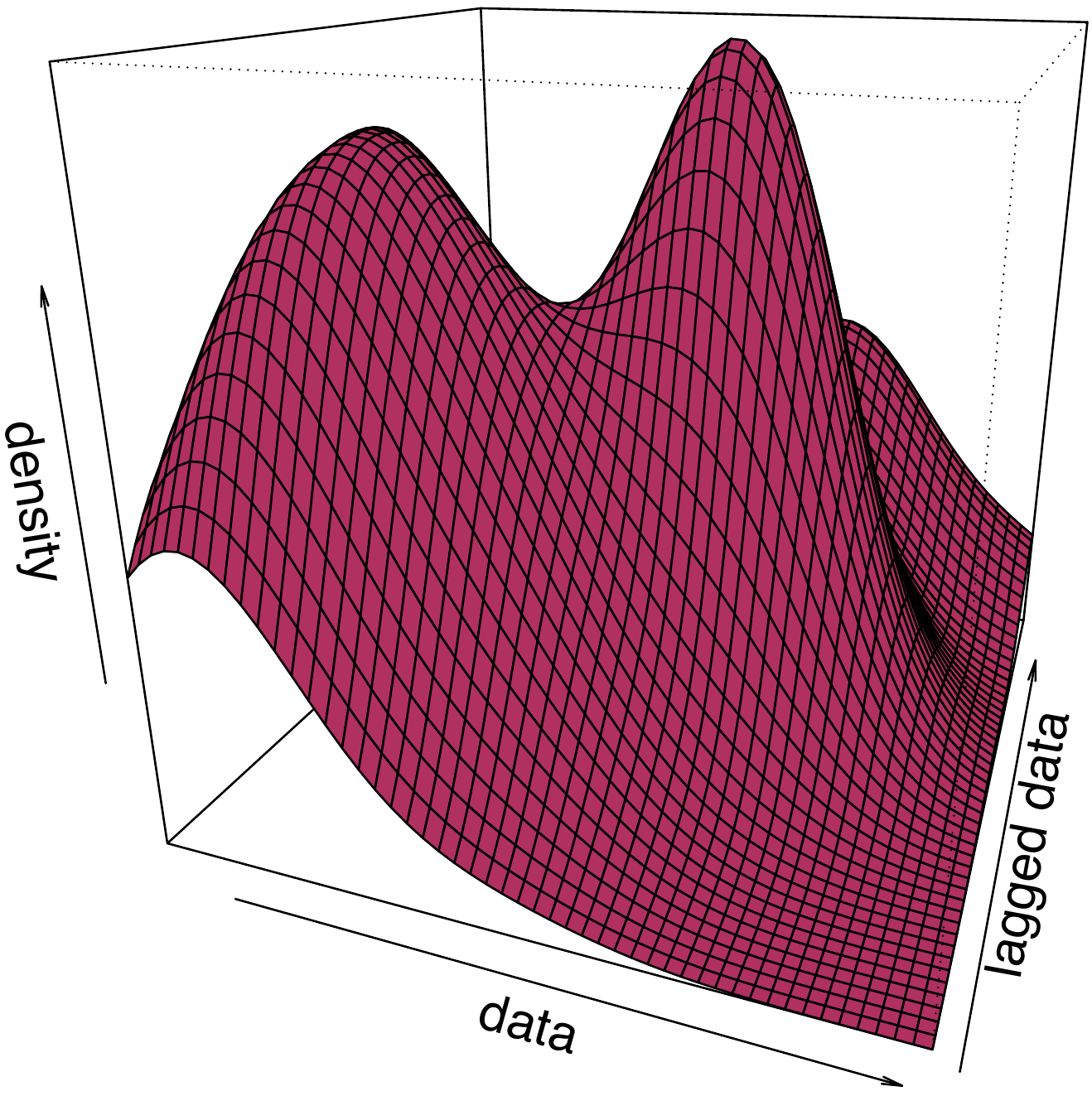} 	\\
		(a) auto-correlation & (b) cross-correlation	
	\end{tabular}
	\caption{Space-time relation between speed measurements. Left panel (a) shows empirical density estimation for the $(s^{10}_n, s^{10}_{n-8})$ bivariate random variable, where $s_n$ is the speed measured at sensor $10$ at time $n$. Right panel (b) shows  empirical density estimation for the $(s^{10}_n, s^{20}_{n-8})$ bivariate random variable.}
	\label{fig:acf}
\end{figure}

As described above, the traffic congestion usually originates at a specific bottlenecks on the network. Therefore, given a bottleneck, our forecast predicts how fast it will propagate on the network. Figure \ref{fig:acf}, shows that the the spatial-temporal relations in traffic data  is non linear.
We now show how to build a deep learning predictor that can capture the nonlinear nature, as well as spatial-temporal patterns in traffic flows.

\subsection{Predictor Selection}
Our deep learning model will estimate  an input-output map, $\pred{t+h}{t} = \hat{y}_{w,b}(x^t)$, where $(w,b)$  index weights and parameters and $x^t$ is the vector of measurements. Our prior assumption is that to predict traffic at a location  we need to use recent measurements from all other sensors on the network.  We use previous 12 measurements from each sensor that corresponds to one hour period.

%

One caveat is that it is computationally prohibitive to use data from every road segment to develop a forecast for a given location and there is some locality in the casual relations between congestion patterns on different road segments. For example, it is unreasonable to assume that a congested arterial in a central business district is related to congestion in a remote suburb, sixty miles away. Thus, it might appear logical to select neighbor road segments as predictors. However, it leads to a large number of unnecessary predictors. For example, congestion in one direction (i.e., towards the business district) does not mean there will be congestion in the opposite direction, which leads to the possibility of using topological neighbors as predictors. The caveat is that by using topological neighbors, it is possible not to include important predictors. For example, when an arterial road is an alternative to a highway, those roads will be strongly related, and will not be topological neighbors. 

\begin{figure}[H]
	\begin{center}
		\includegraphics[width=0.8\linewidth]{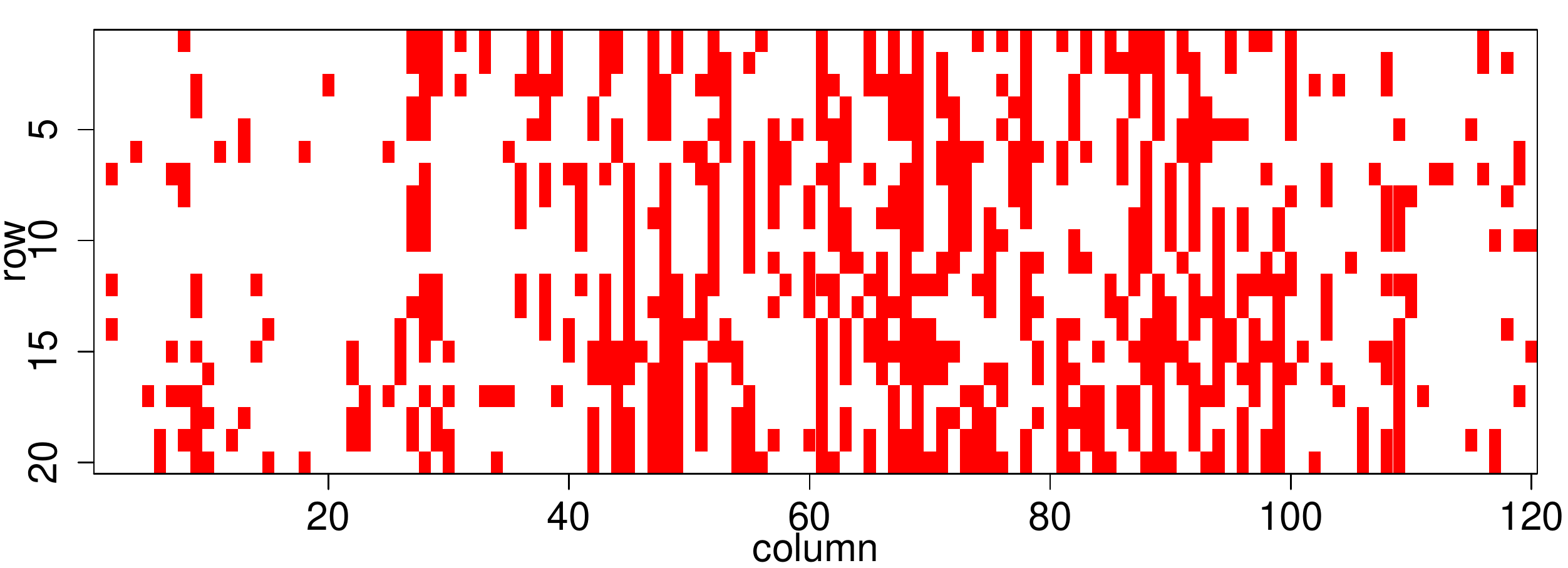}
	\end{center}
	\caption{Sparsity patterns of the coefficient matrix found by least angle regression (LARS)}
	\label{fig:lasso-a}
\end{figure}

Methods of regularization on the other hand provide the researcher  with an automated approach to select predictors. Least angle regression (LARS) is used to fit the $\ell_1$ regularized loss function (LASSO) and to find a sparse linear model. Figure \ref{fig:lasso-a} shows the sparsity pattern of the  resulting coefficient matrix.

\begin{figure}[H]
	\includegraphics[width=1\textwidth]{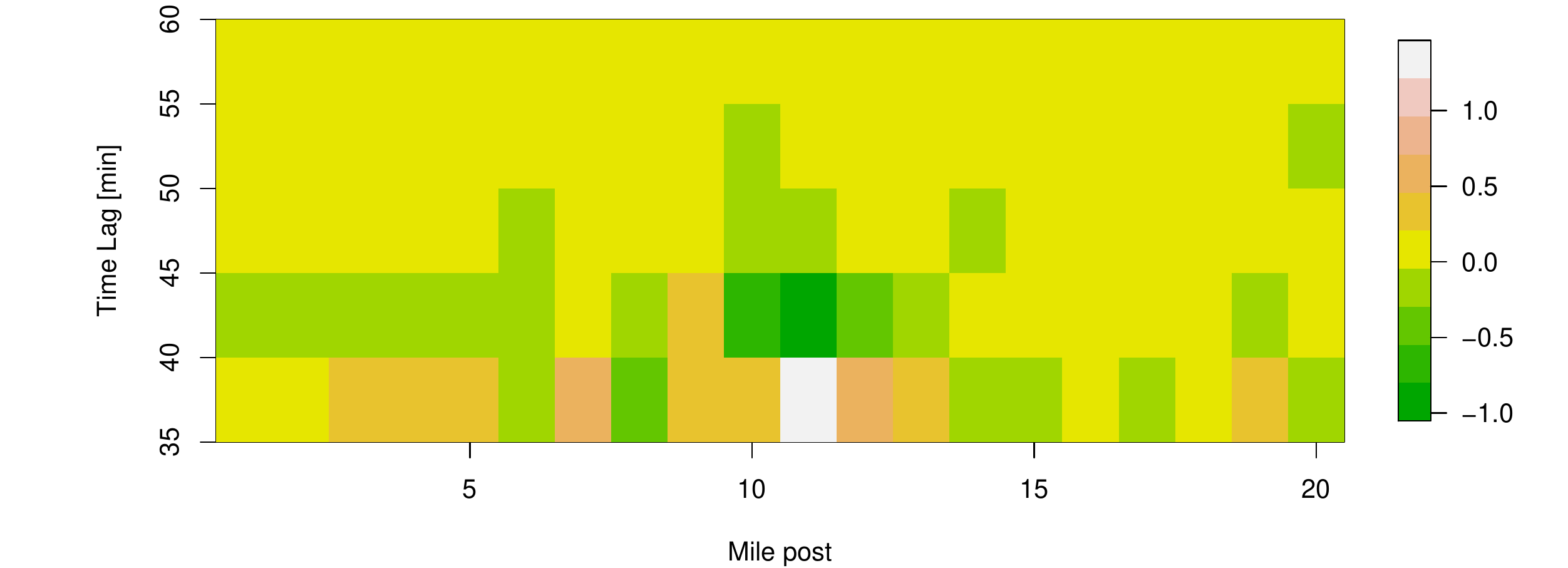}
	\caption{Values of coefficients of the fitted linear model for predicting sensor 11.  }
	\label{fig:betas}
\end{figure}

Figure~\ref{fig:betas} shows the magnitude of the coefficients of the linear model for sensor 11. There are 120 predictors used for each of the locations that correspond to 6 lagged measurements from 20 sensors. We see that the largest coefficient is the one that corresponds to the most recent measurement from the sensor itself (white-colored rectangle). We also see that the model does assign the largest values to variables that are close to the modeled variable in time and space. Most of the weight will be on the most recent measurement (lag of 35 minutes). The previous measurement, which corresponds to a lag of 40 minutes, has negative weight. It means that the weighted difference between two consecutive measurements is used as a predictor. Intuitively, it means that the change in speed is the predictor rather than the absolute value of speed. In a time series context, negative weights correspond to cyclic behavior in the data, see~\cite{shumway2010time}. 

Another way is to find a sparse  neural network model is to apply a dropout operation. Suppose that we have an $\ell_2$ objective
$$
\argmin_{w,b} \; \Vert \y - \hat{\y}_{w,b} (\myx) \Vert_2^2.
$$
Due to the composite nature of the predictor, we can calculate derivative information $ \nabla_{w,b} \; l( y , \hat{y}_{w,b} ( x ) ) $ using the chain rule via procedure known as backpropagation.

To perform model or variable selection, suppose that we dropout any input dimension in $x$ with probability $p$. This replaces the input by $ D \star x $ where $ \star$ denotes element-wise products and $D$ is a matrix of $ Ber(p)$ random variables. Marginalize over the randomness, we have a new objective
$$
(\hat{w}, \hat{b}) \in \argmin_{w,b} \; E_{ D \sim {\rm Ber} (p) }\left( \Vert \y - \hat{\y}_{D \star w,b} (\myx) \Vert_2^2\right),
$$
which is equivalent to
$$
(\hat{w}, \hat{b}) \in\argmin_{w,b} \; \Vert \y - \hat{\y}_{w,b} (\myx) \Vert_2^2 + p(1-p) \Vert \Gamma w \Vert^2,
$$
where $ \Gamma = ( {\rm diag} ( X^\top X) )^{\frac{1}{2}} $ and $X$ is the matrix of observed features. Therefore, the objective function dropout is equivalent to a Bayes ridge regression with a $g$-prior \cite{george2000variable}.

The end model has minimal deviance for a validation data set. We used data from 180 days to fit and test the model, we used the first half of the selected days for training and the remaining half for testing the models.

\subsection{Chicago Highway Case Study Results}
Having motivated our modeling approach and described general traffic flow patterns, we now evaluate predictive power of sparse linear vector autoregressive (VAR) and deep learning models. Using loop detector data from 21 sensors installed on Chicago's Interstate highway I-55 measure in  the year of 2013. They cover a 13-mile stretch of a highway that connects southwest suburbs to  Chicago's downtown. In our empirical study,  we predict traffic flow speed at the location of senor 11, which is in the middle of the 13-mile stretch. It is located right before Cicero Avenue exit from I-55. We treated missing data by doing interpolation on space, i.e. the missing speed measurement $s_{it}$ for sensor $i$ at time $t$ will be amputated using $(s_{i-1t} + s_{i+1t})/2$. Data from  days when the entire sensor network was down were excluded from the analysis. We also excluded public holidays and weekend days.

We compare the performance of the deep learning (DL) model with sparse linear vector autoregressive (VAR), combined with several data pre-filtering techniques, namely median filtering with a window size of 8 measurements (M8) and  trend filtering with $\lambda = 15$ (TF15). We also tested performance of the sparse linear model, identified via regularization. We estimate the percent of variance explained by model, and mean squared error  (MSE), which measures average of the deviations between measurements and model predictions. To train both models we selected data for 90 days in 2013. We further  selected another 90 days for testing data set. We calculated $R^2$ and MSE for both in-sample (IS) data and out-of-sample (OS) data. Those metrics are shown in Table~\ref{tab:results}.
\begin{table}[H]
	\centering
	\fbox{%
	\begin{tabular}{l|ccccccc}
	&DLL  & DLM8L & DLM8  & DLTF15L    & DLTF15 		 & VARM8L & VARTF15L \\ \hline
	IS MSE   &  13.58       & 7.7                      &  10.62    &  12.55 		  &12.59     &  8.47				&15\\      
	IS $R^2$ &0.72        & 0.83                    &  0.76      &  0.75			 & 0.75  	 &  0.81				&0.7\\ 
	OS MSE     & 13.9     & 8.0                      &  9. 5       &  11.17		   &  12.34	  &  8.78				 &15.35\\ 
	OS $R^2$   &  0.75  & 0.85                   &  0.82      &  0.81			   &  0.79	    &  0.83				  &0.74\\ 
	\end{tabular}} 	
\caption{In sample and out-of-sample metrics for different models. The abbreviations for column headers are as follows:  DL = deep learning, VAR = linear model, M8 = media filter preprocessing, TF15 = trend filter preprocessing and  L = sparse estimator (lasso). The abbreviations for row headers are as follows: IS = in-sample, MSE = mean squared error and OS  = out-of-sample. }
\label{tab:results}
\end{table}

\begin{figure}[!h]
\hspace*{-.5in}
\centering
\makebox{
\begin{tabular}{l|ccc}
	model & bears game  & weather day  & normal day \\\hline
	DLM8L & \includegraphics[width=0.33\linewidth]{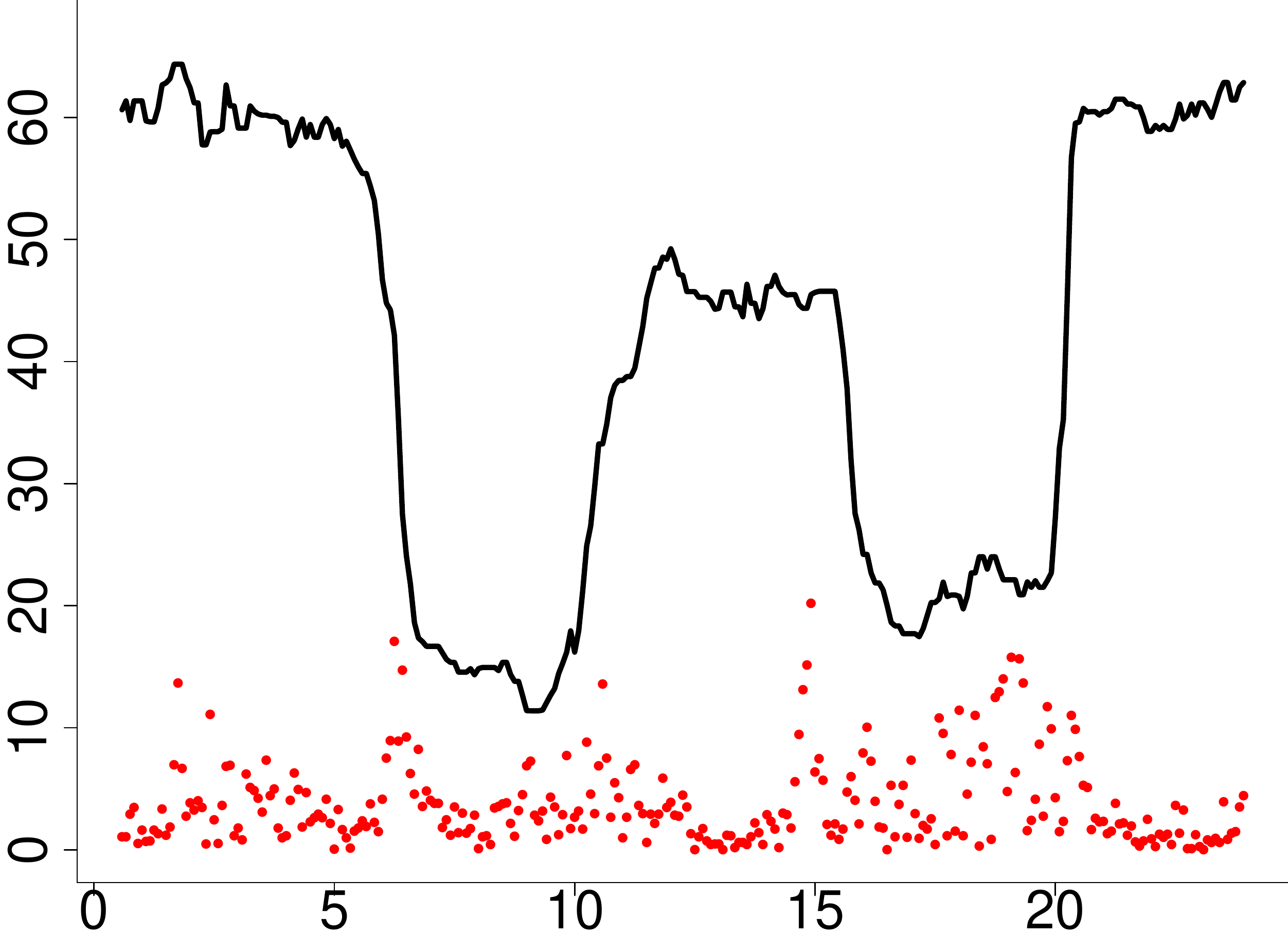} & \includegraphics[width=0.33\linewidth]{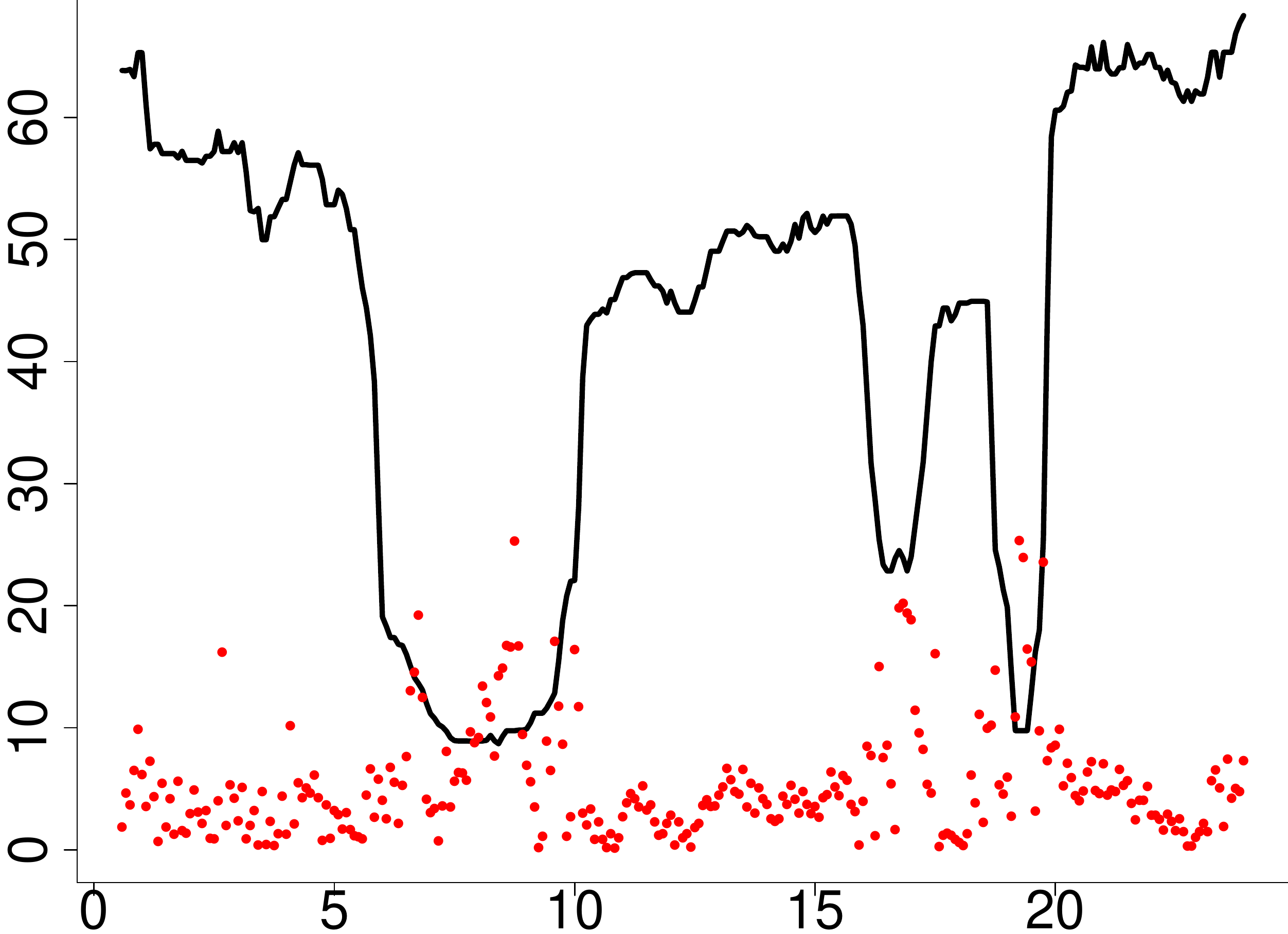} &
	\includegraphics[width=0.33\linewidth]{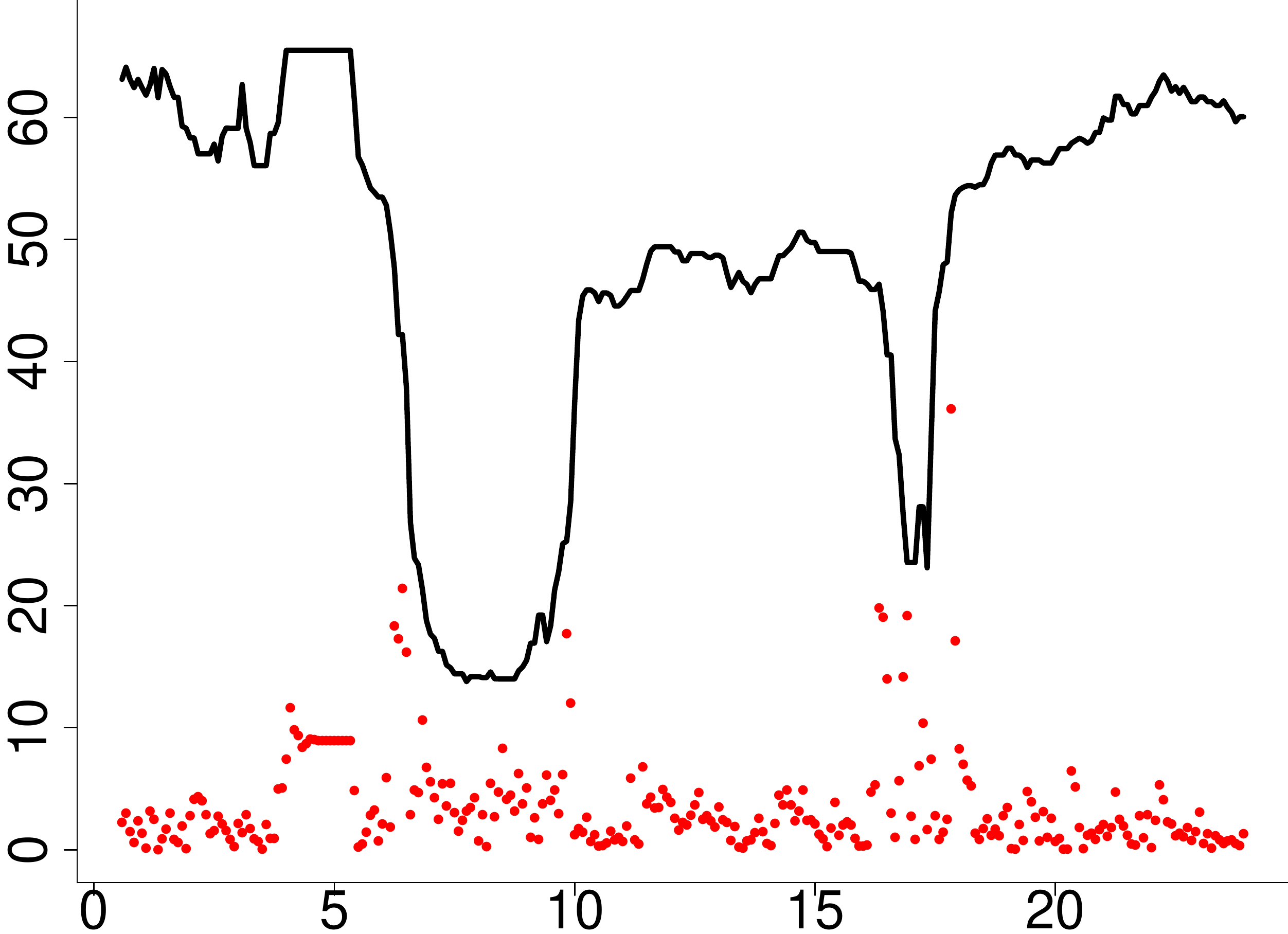}\\
	VARM8L & \includegraphics[width=0.33\linewidth]{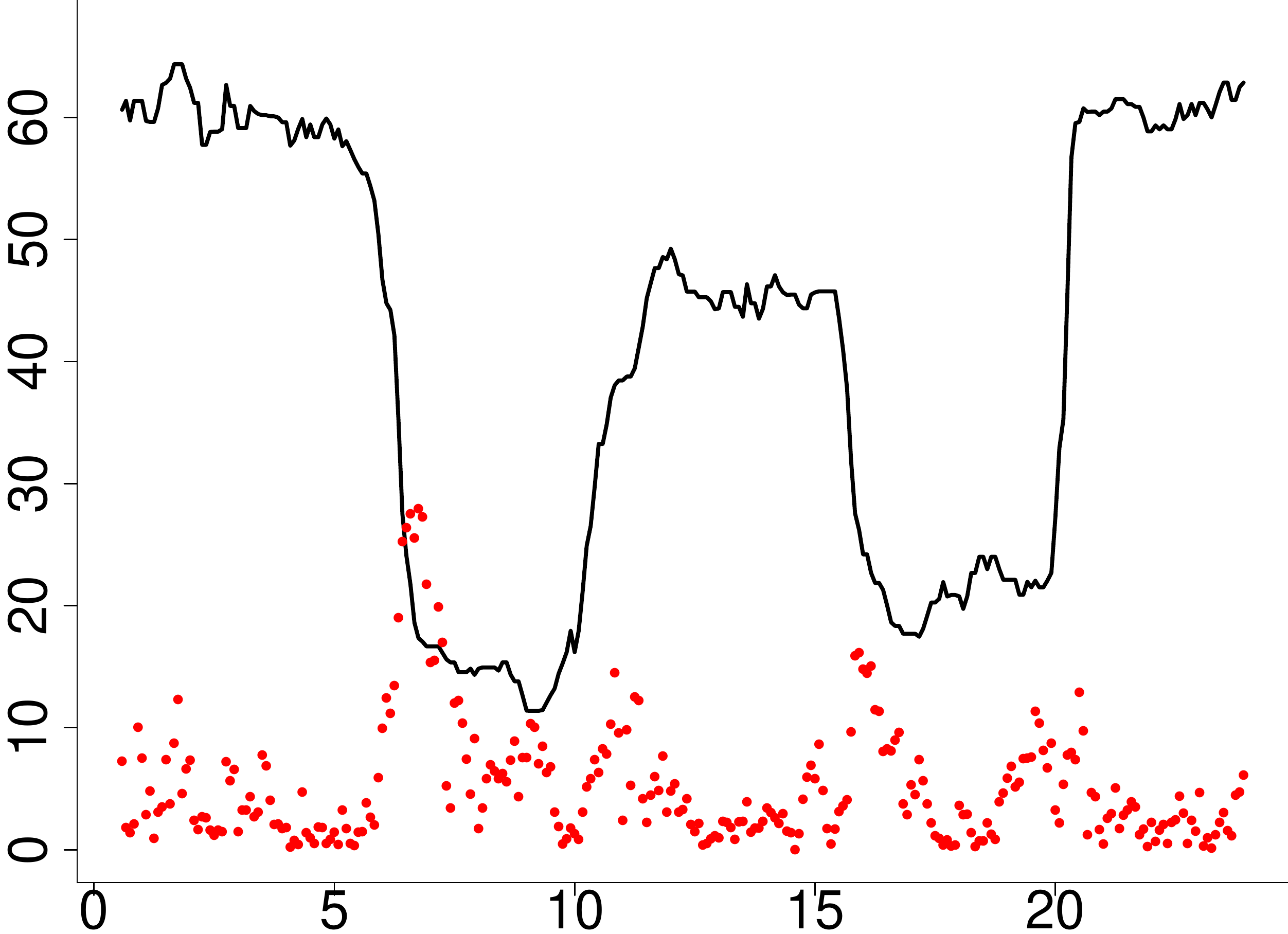} & \includegraphics[width=0.33\linewidth]{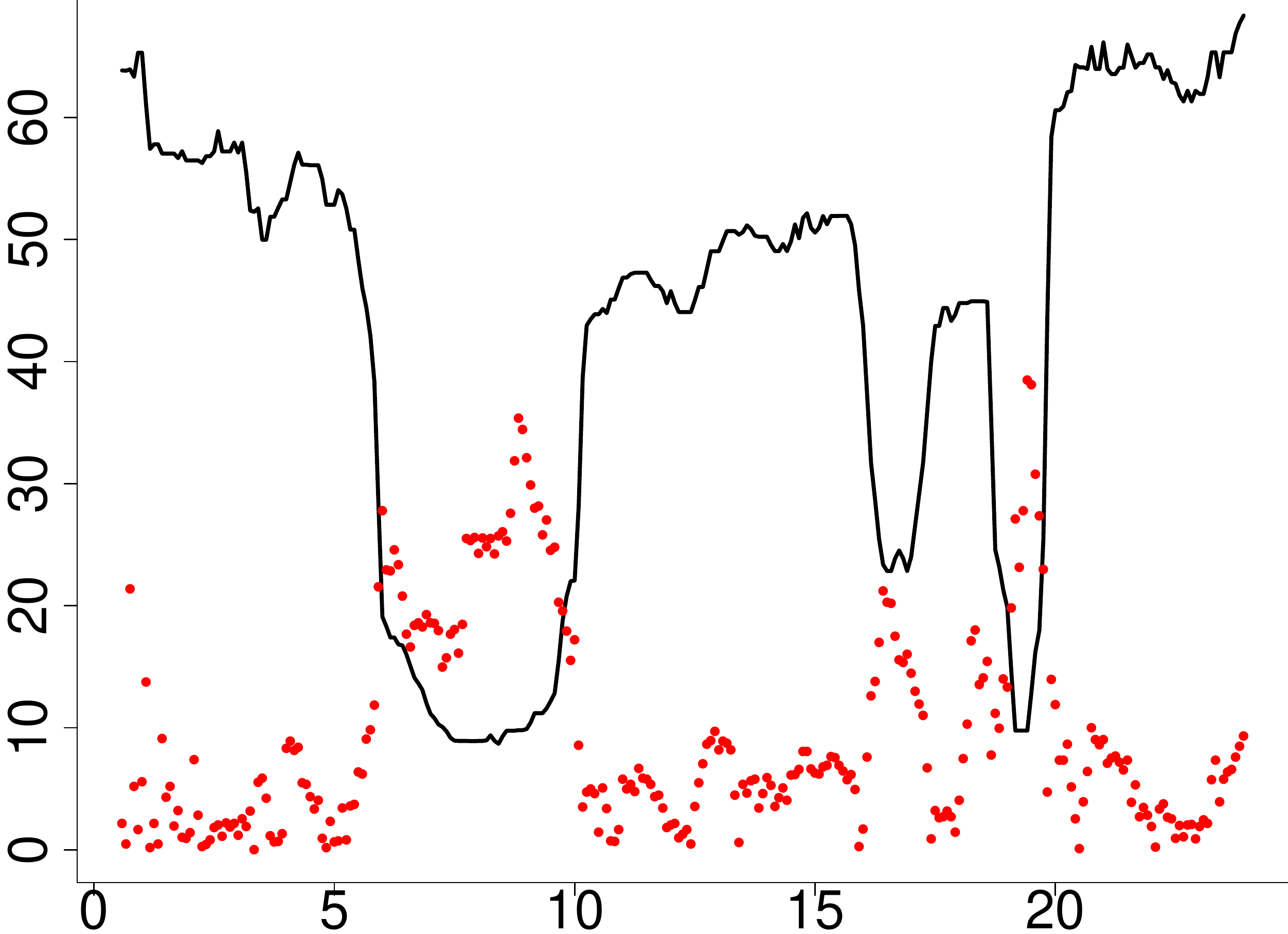} &
	\includegraphics[width=0.33\linewidth]{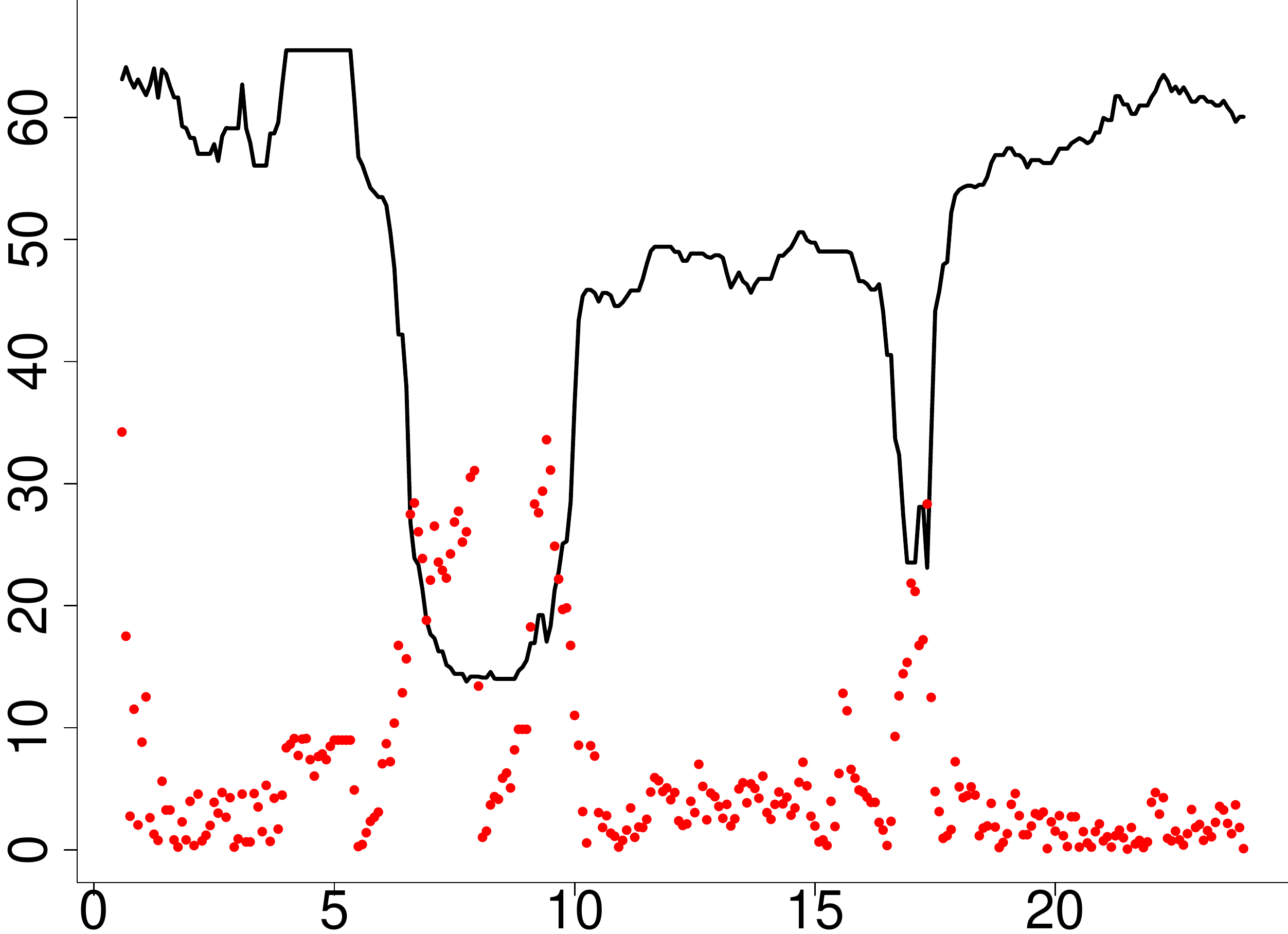} \\

\end{tabular}}
\caption{Comparison of the residuals forecasts over time. On all plots \textbf{black} solid line is the measured data (cross-section speed), {\color{red} \textbf{red}} dots are absolute values of residuals from our model's forty minute forecast. First column compares models for data from Thursday October 10, 2013, the day when Chicago Bears team played New York Giants. The game starts at 7pm and lead to an unusual congestion starting at around 4pm. Second column compares models for data from Wednesday December 11, 2013, the day of light snow. The snow leads to heavier congestion during both, the morning and  evening rush hours. Third column compares models for data from  Monday October 7, 2013. There were no special events, accidents or inclined weather conditions on this day.}
\label{fig:model_res}
\end{figure}

The performance of the model is not uniform throughout the day. Figure \ref{fig:model_res} shows the absolute values of the residuals (red circles) against the measured data (black line). Highest residuals are observe at when traffic flow regime changes from one to another. On a normal day large errors are observed at around 6am, when regime changes from free flow to congestion and at around 10am, right before it starts to recover back to free flow. 

Sparse deep learning combined with the median filter pre-processing (DLM8L) shows the best overall performance on the out-of-sample data. 

\begin{figure}[!h]
\hspace*{-.5in}
\centering
\makebox{
\begin{tabular}{l|ccc}
	model & bears game  & weather day  & normal day \\\hline
	DLM8L & \includegraphics[width=0.33\linewidth]{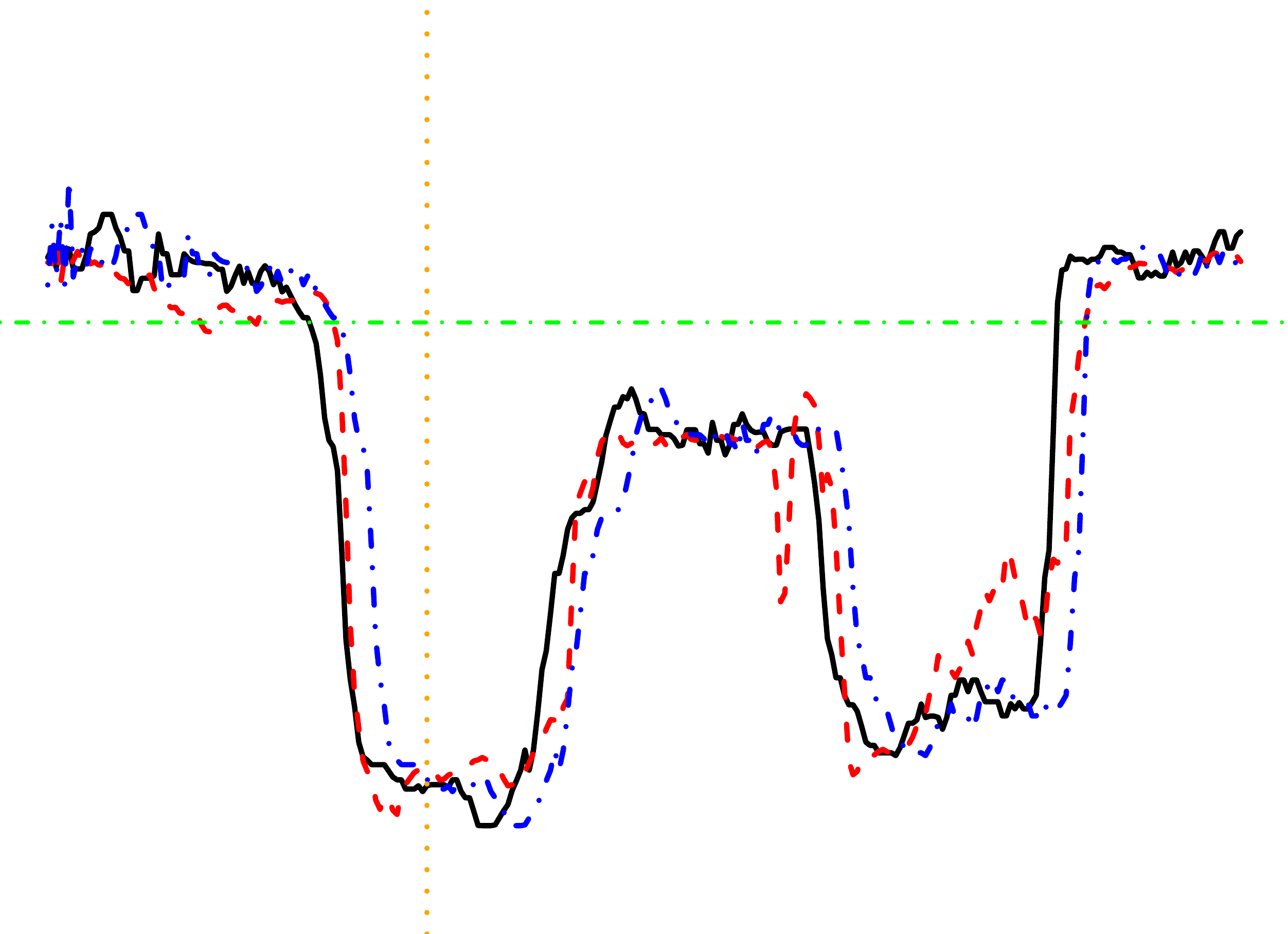} & \includegraphics[width=0.33\linewidth]{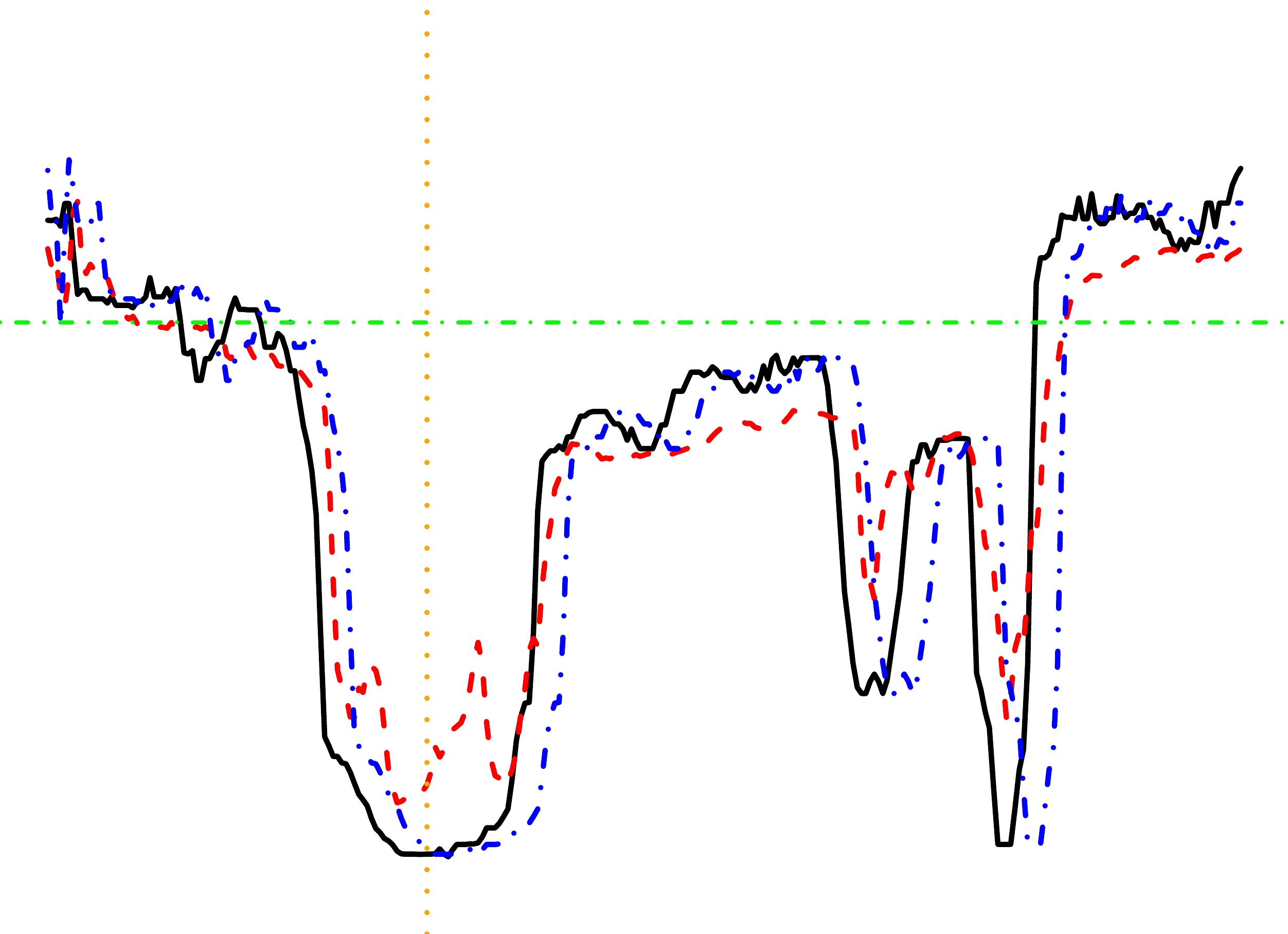} &
	\includegraphics[width=0.33\linewidth]{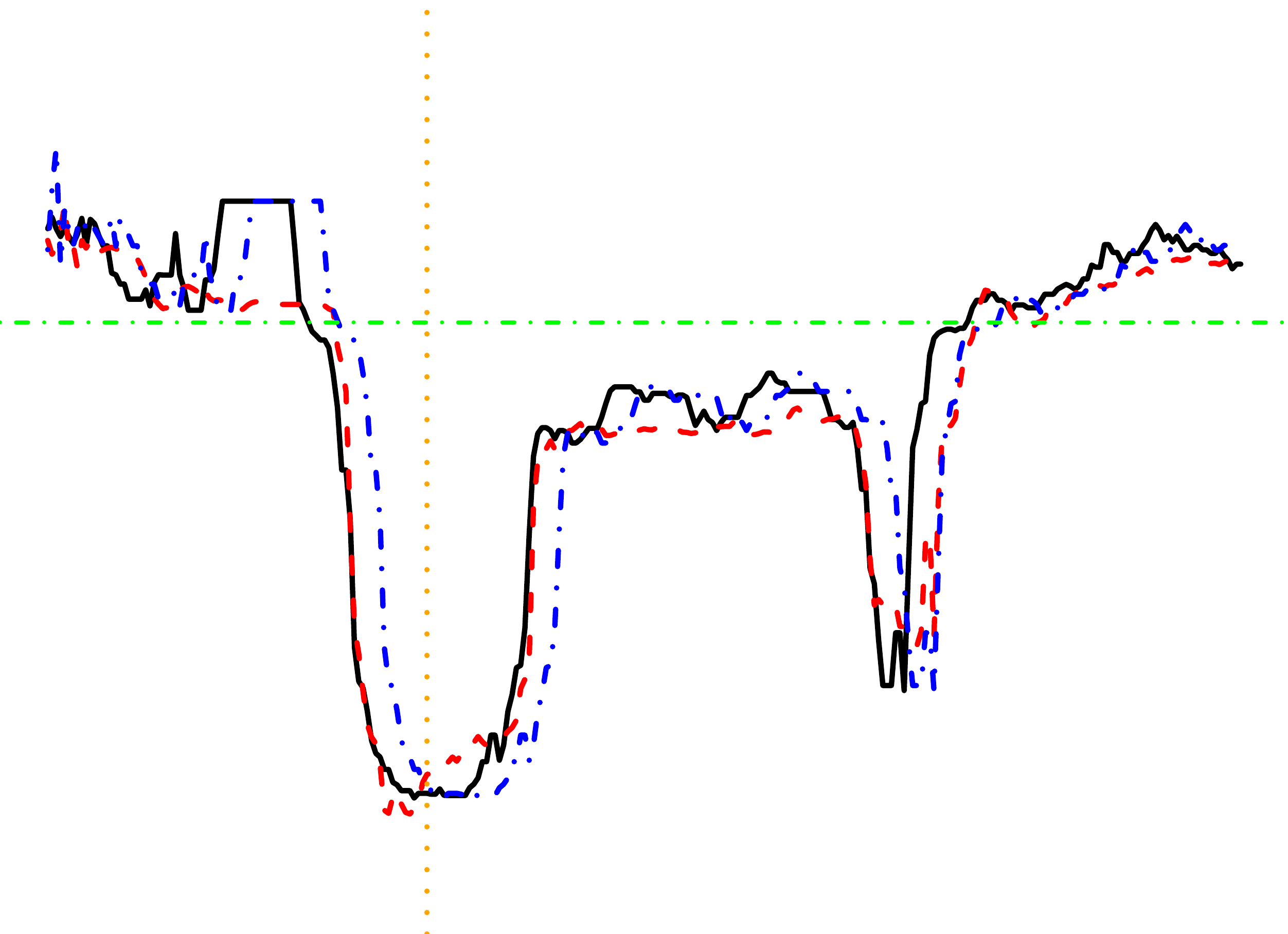}\\
	VARM8L & \includegraphics[width=0.33\linewidth]{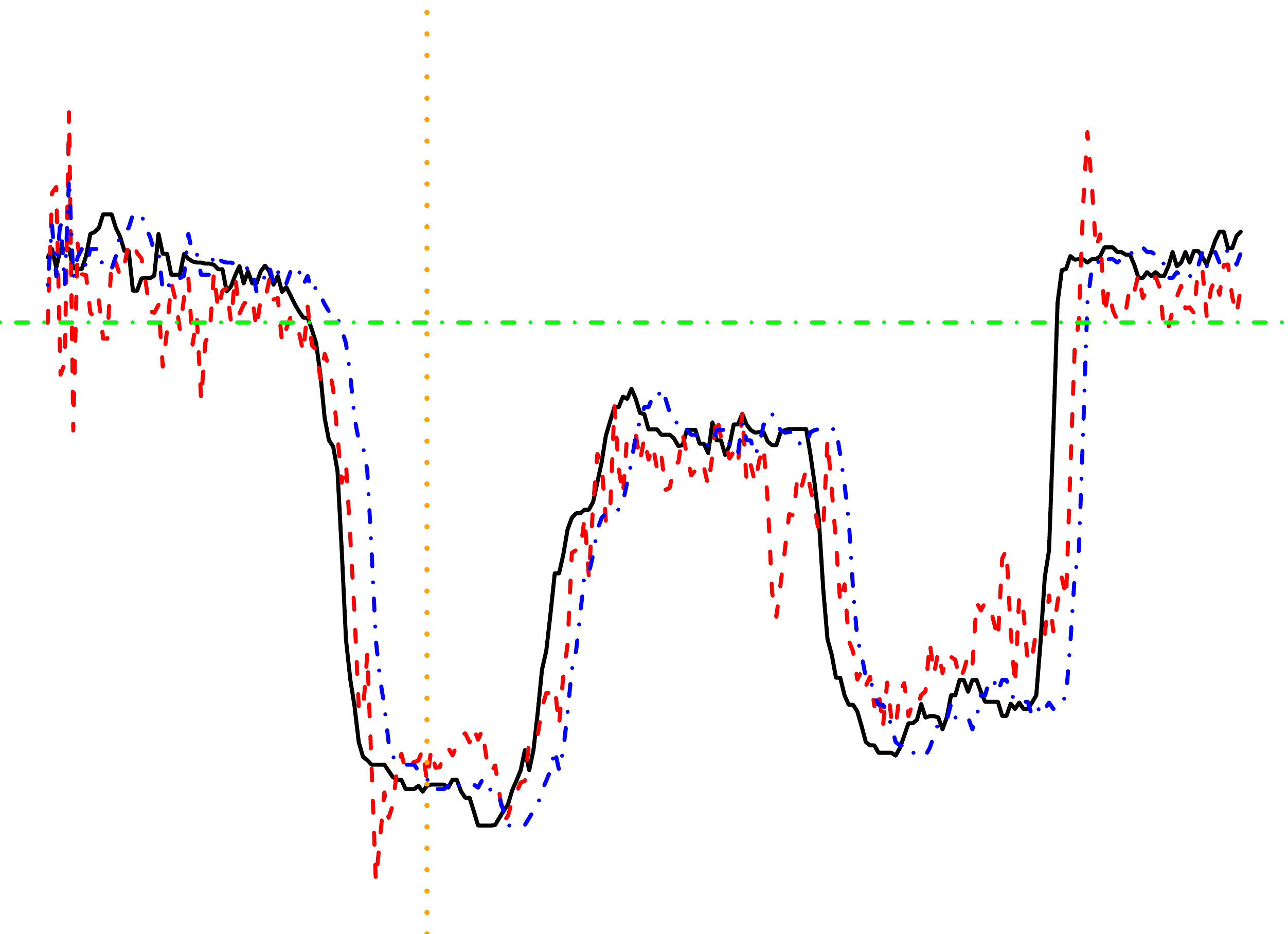} & \includegraphics[width=0.33\linewidth]{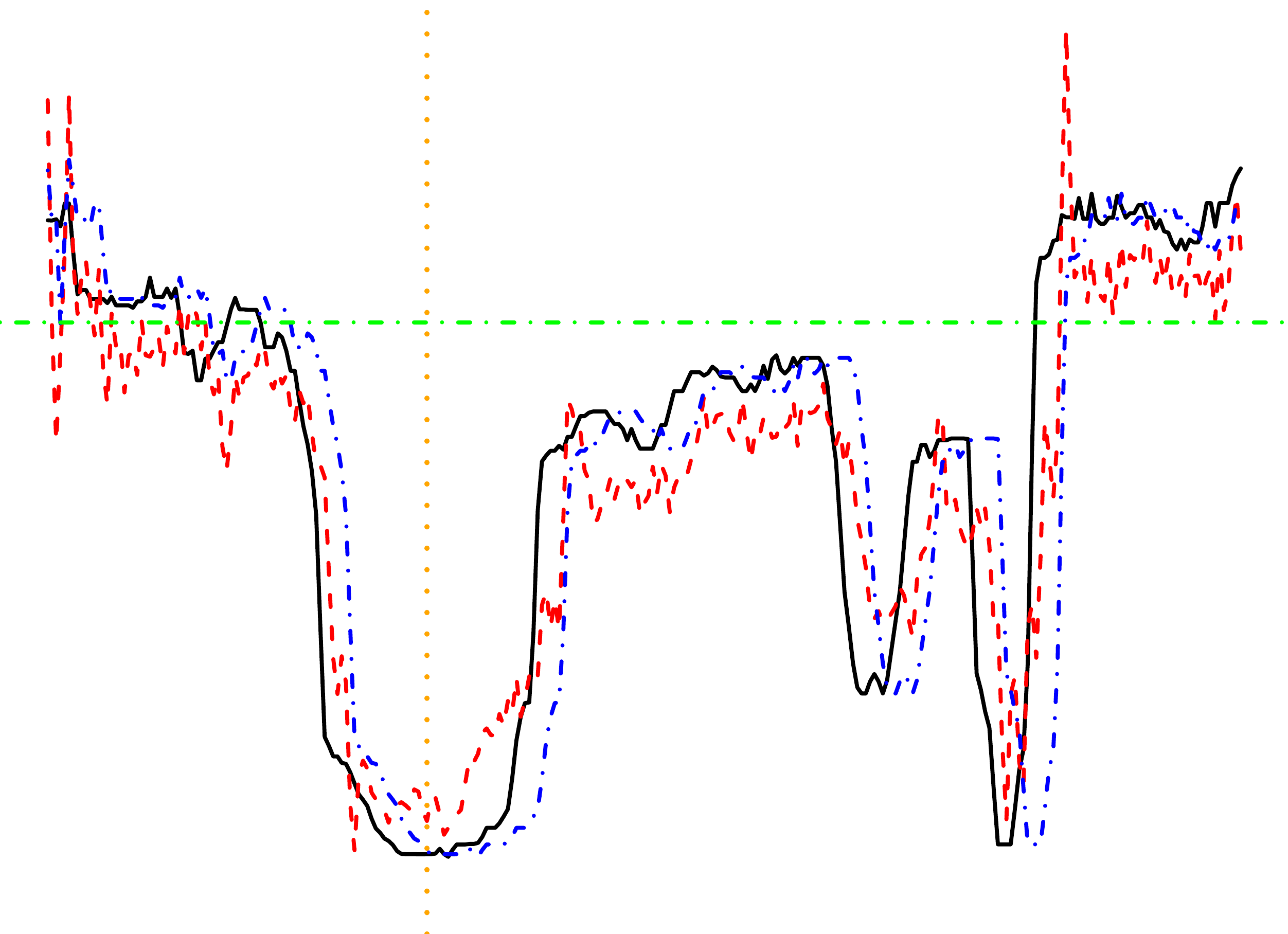} &
	\includegraphics[width=0.33\linewidth]{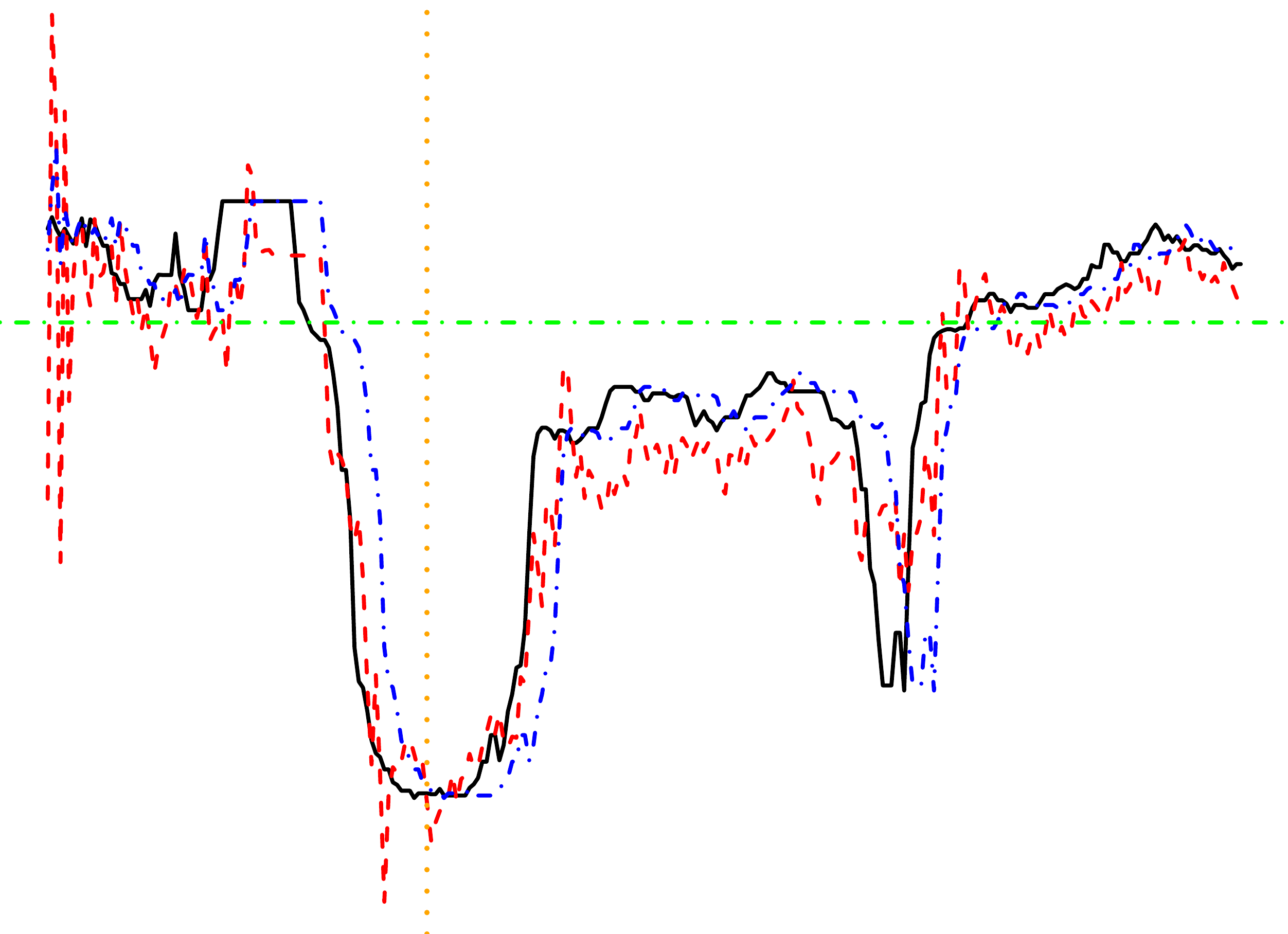} \\
	DLL & \includegraphics[width=0.33\linewidth]{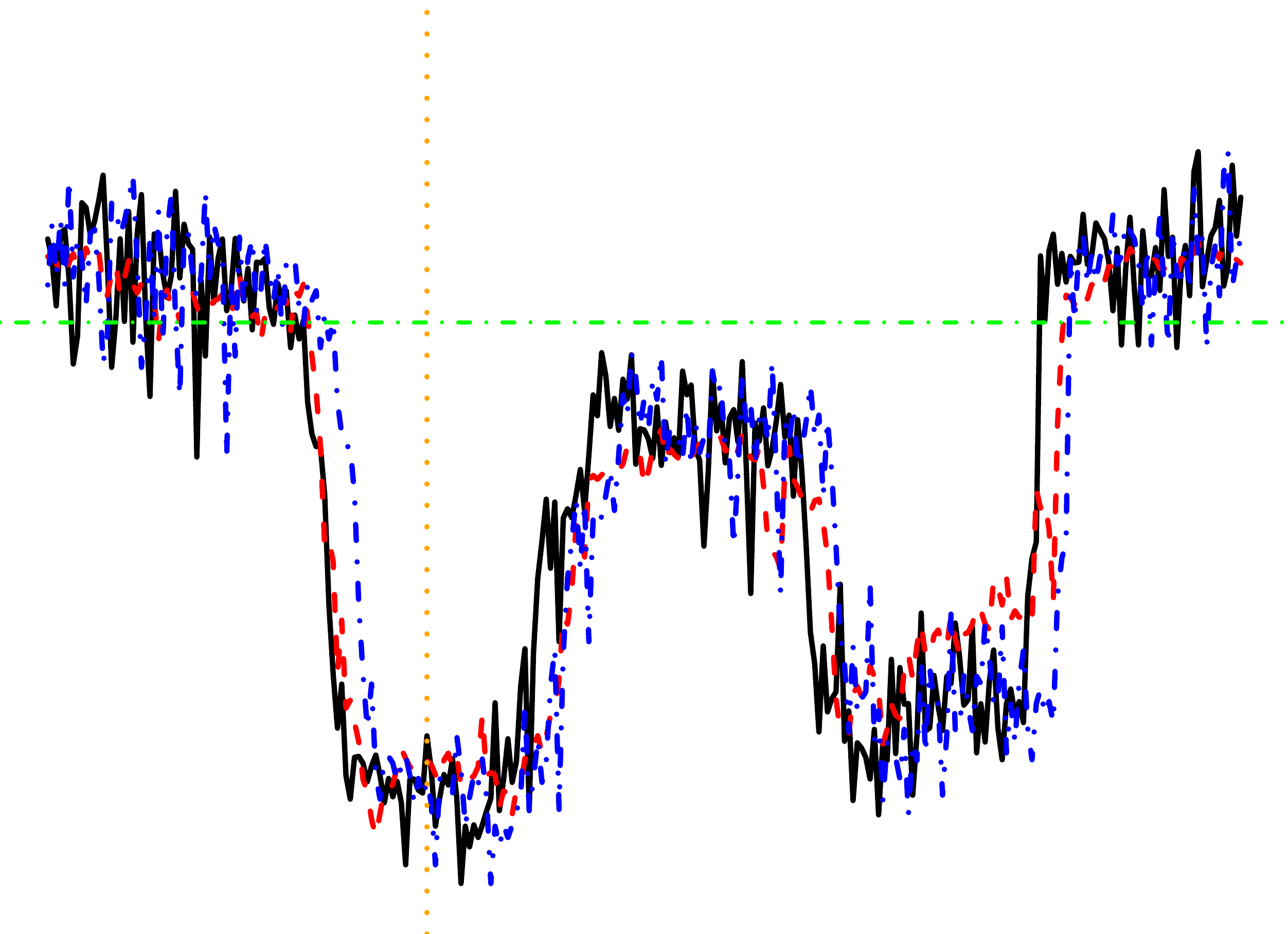} & \includegraphics[width=0.33\linewidth]{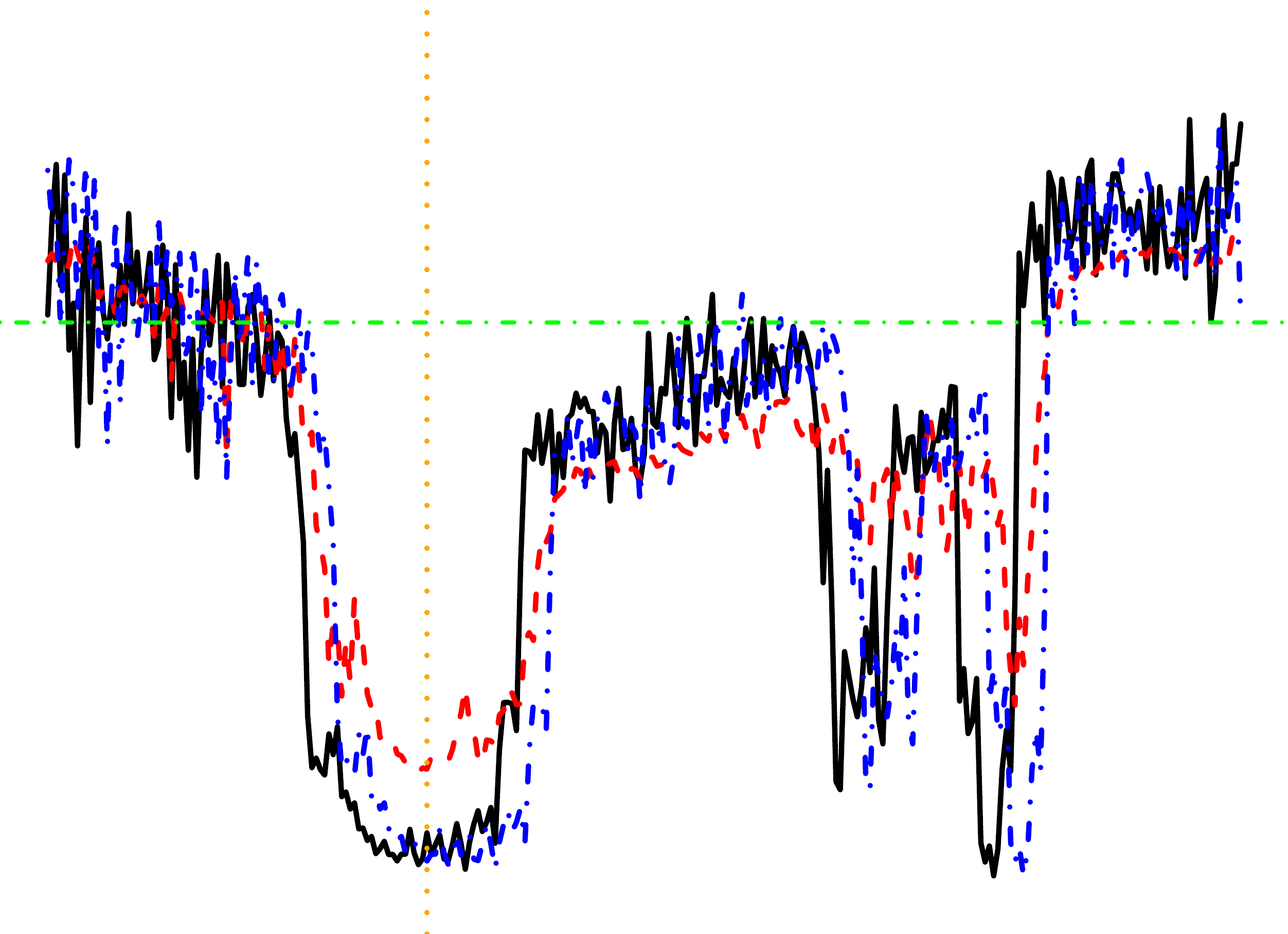} &
		\includegraphics[width=0.33\linewidth]{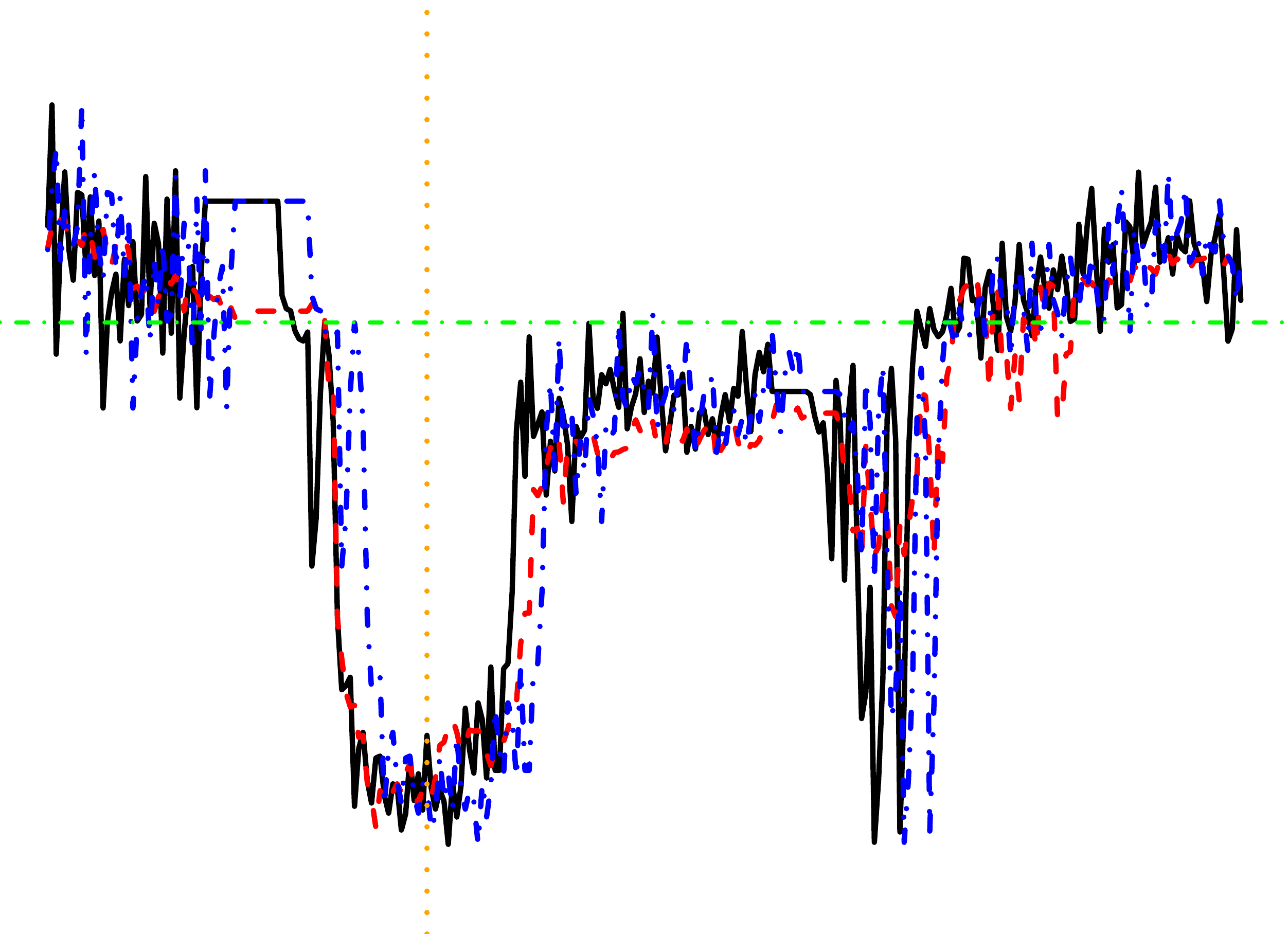}\\				
\end{tabular}}
\caption{Comparison of the forecasts. On all plots \textbf{black} solid line is the measured data (cross-section speed), {\color{red} \textbf{red}} dashed line is our model's forty minute forecast and dashed {\color{blue} \textbf{blue}} line is naive forecast. Green dashed horizontal line is the speed limit (55 mi/h) and vertical orange line is the morning peak hour (8am).  First column compares models for data from Thursday October 10, 2013, the day when Chicago Bears team played New York Giants. The game starts at 7pm and lead to an unusual congestion starting at around 4pm. Second column compares models for data from Wednesday December 11, 2013, the day of light snow. The snow leads to heavier congestion during both, the morning and  evening rush hours. Third column compares models for data from  Monday October 7, 2013. There were no special events, accidents or inclined weather conditions on this day.}
\label{fig:model}
\end{figure}

Figure~\ref{fig:model} shows performance of both vector auto-regressive and deep learning models for normal day, special event day (Chicago Bears football game) and poor weather day (snow day). We compare our models against the naive constant filter, i.e forecast speed is the same as the current speed. The naive forecast is used by travelers when making route choices before departure. We achieve this by looking at current traffic conditions and assuming those will hold throughout the duration of the planned trip. 

Both deep learning (DL) and  vector auto-regressive (VAR) models accurately predict the morning rush hour congestion on a normal day. However, the vector auto-regressive model mis-predicts congestion during evening rush hour. At the same time, deep learning model does predict breakdown accurately but miss-estimates the time of recovery. Both deep learning and linear model outperform naive forecasting, when combined with data pre-processing. However, when unfiltered data is used to fit deep learning combined with sparse linear estimator (DLL) model, their predictive power degrades and were out-performed by a naive forecast. Thus, showing the importance of using filtered data to develop forecasts.

Typically, congestion starts at a bottleneck location and propagates downstream. However, traffic slows down at several locations at the same time during the snow. Thus, for all three models, there is a lag between the forecast and real data for the "weather day" case. We think, that adding a weather forecast as a predictor variable will improve forecasts for traffic caused by snow or rain. The vector autoregressive  (VAR) model forecasts lag the data for all three days. 
Our vector autoregressive model shows surprisingly good performance predicting the normal day traffic, that is comparable to the deep learning model forecast. Deep learning predictor, can  produce better predictions for non-recurrent events, as shown for our Bears game and weather day forecasts example. 

\begin{figure}[H]
	\begin{tabular}{cc}
		\includegraphics[width=0.5\textwidth]{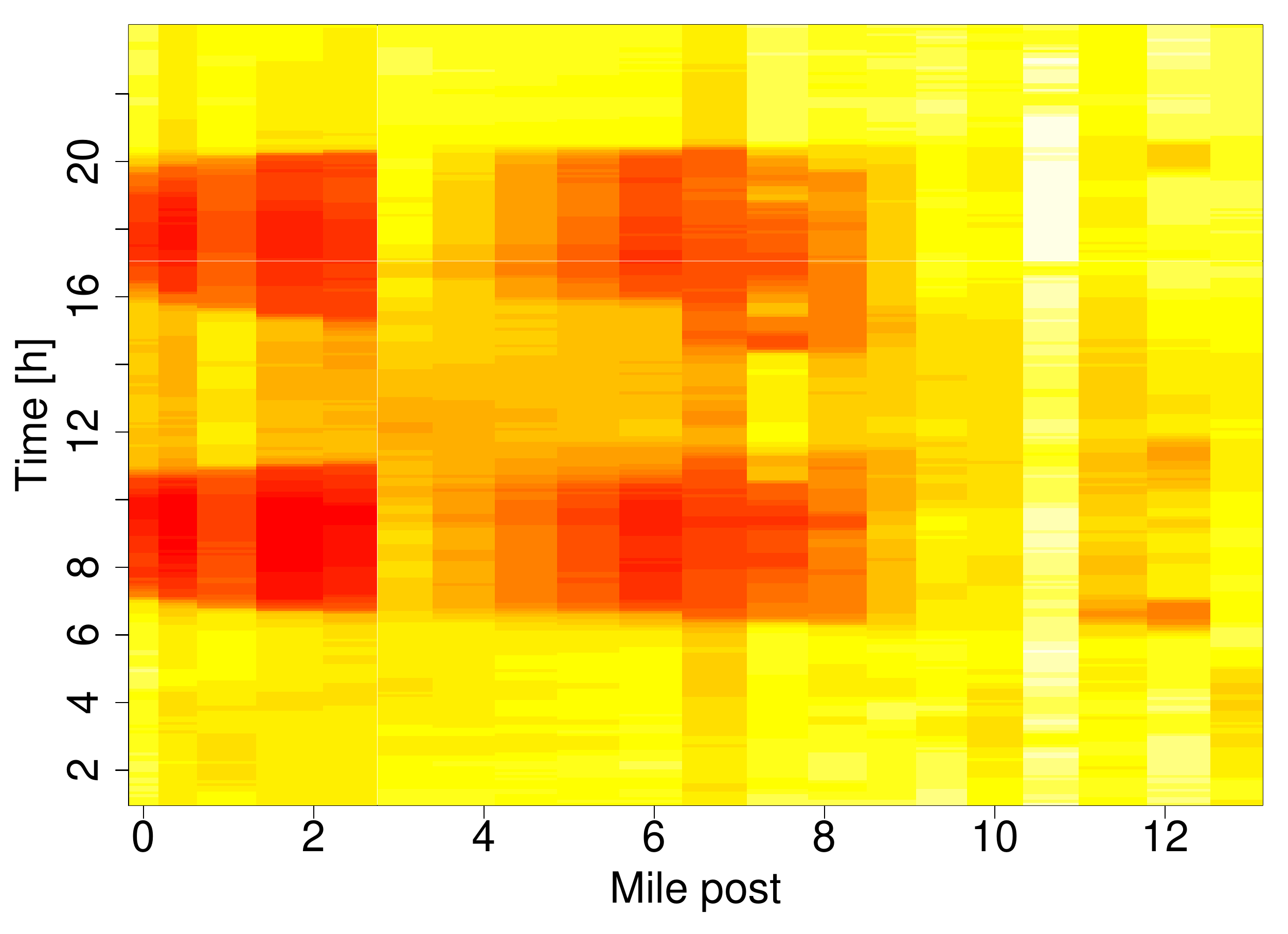}&
		\includegraphics[width=0.5\textwidth]{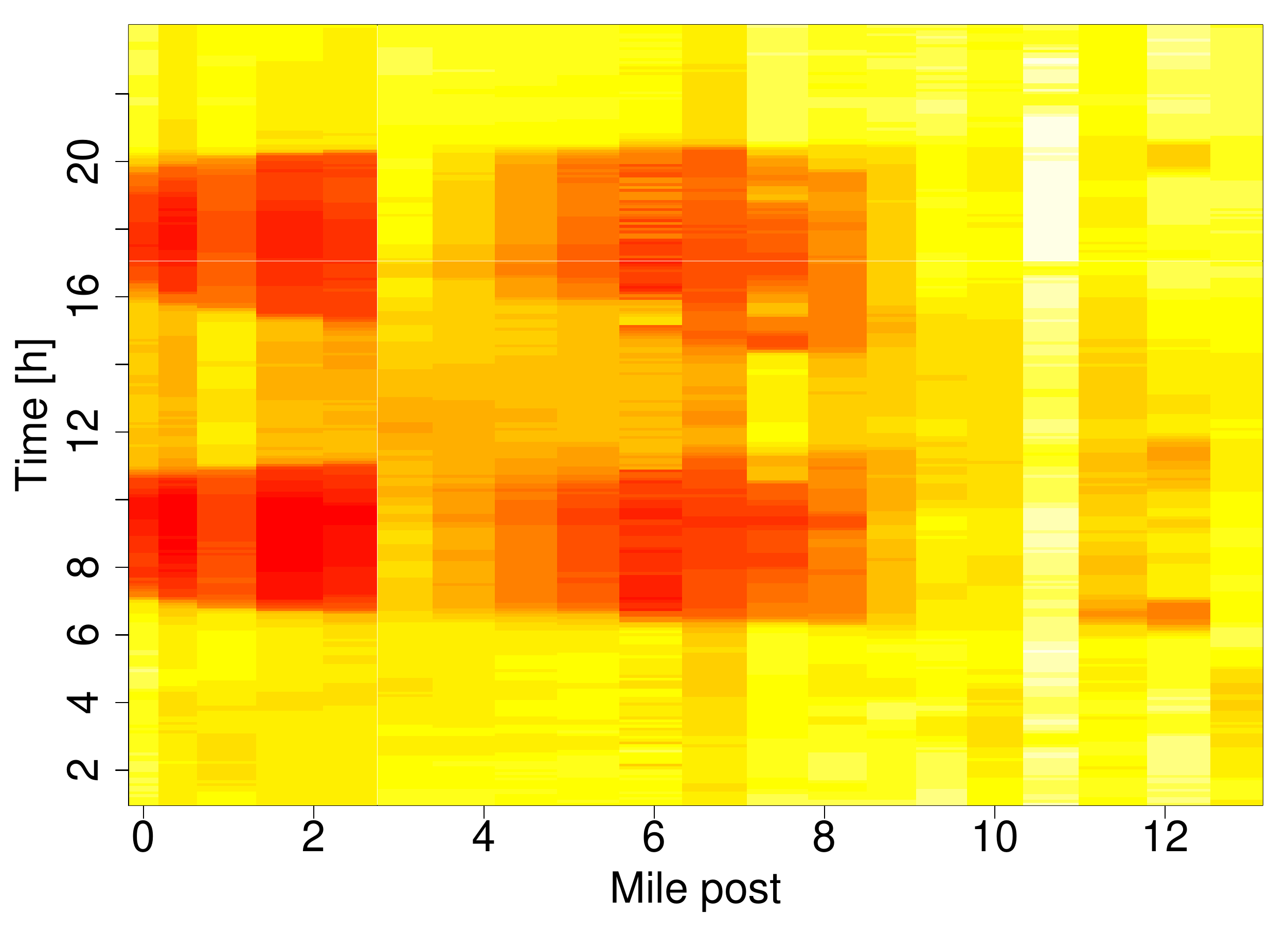}\\
		(a) Original Data & (b) Forecast\\
	\end{tabular}
	\caption{Heat plot for traffic spreads on Thursday October 10, 2013, the day of the Bears game. Right panel (b) was produced by replacing column 11 (mile post 6) of the measured data with forecast for this location.}
	\label{fig:heat_10}
\end{figure}

Another visual way to interpret the results of prediction is via a heat plot. Figure~\ref{fig:heat_10} compares the original data and forecasted data. To produce forecast plot we replaced column 11 of the original data (mile post 6) with the forecast for this specific location.

From Figure~\ref{fig:heat_10} we see that deep learning model properly captures both forward and backward shock wave propagation during morning and evening rush hours. 

\subsection{Residual Diagnostics}
To assess the accuracy of a forecasting model we analyze the residuals, namely the difference between observed value and its forecast $r_i = y_i - \hat{y}_i$. Our goal is to achieve residuals that are uncorrelated with zero mean. In other words, there is no information left in the residuals that can be used to improve the model. Models with the best residuals do not necessarily  have the most forecasting power, out of all possible models, but it is  an important indicator of whether a model uses all available information in the data. 

\begin{figure}[H]
\begin{tabular}{cccc}
	\rotatebox[origin=lt]{90}{VARM8}&
	\includegraphics[width=0.3\textwidth]{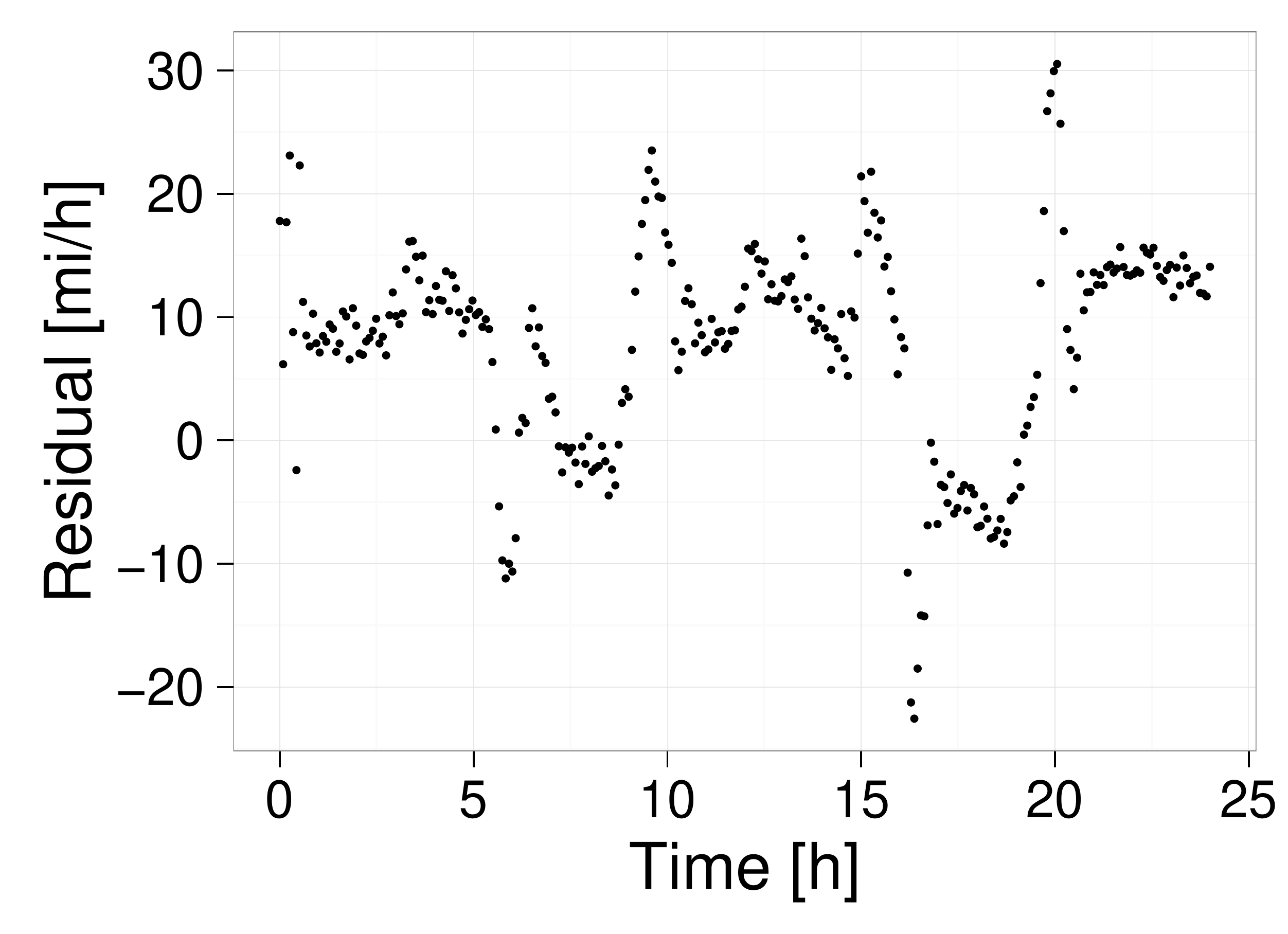}&
	\includegraphics[width=0.3\textwidth]{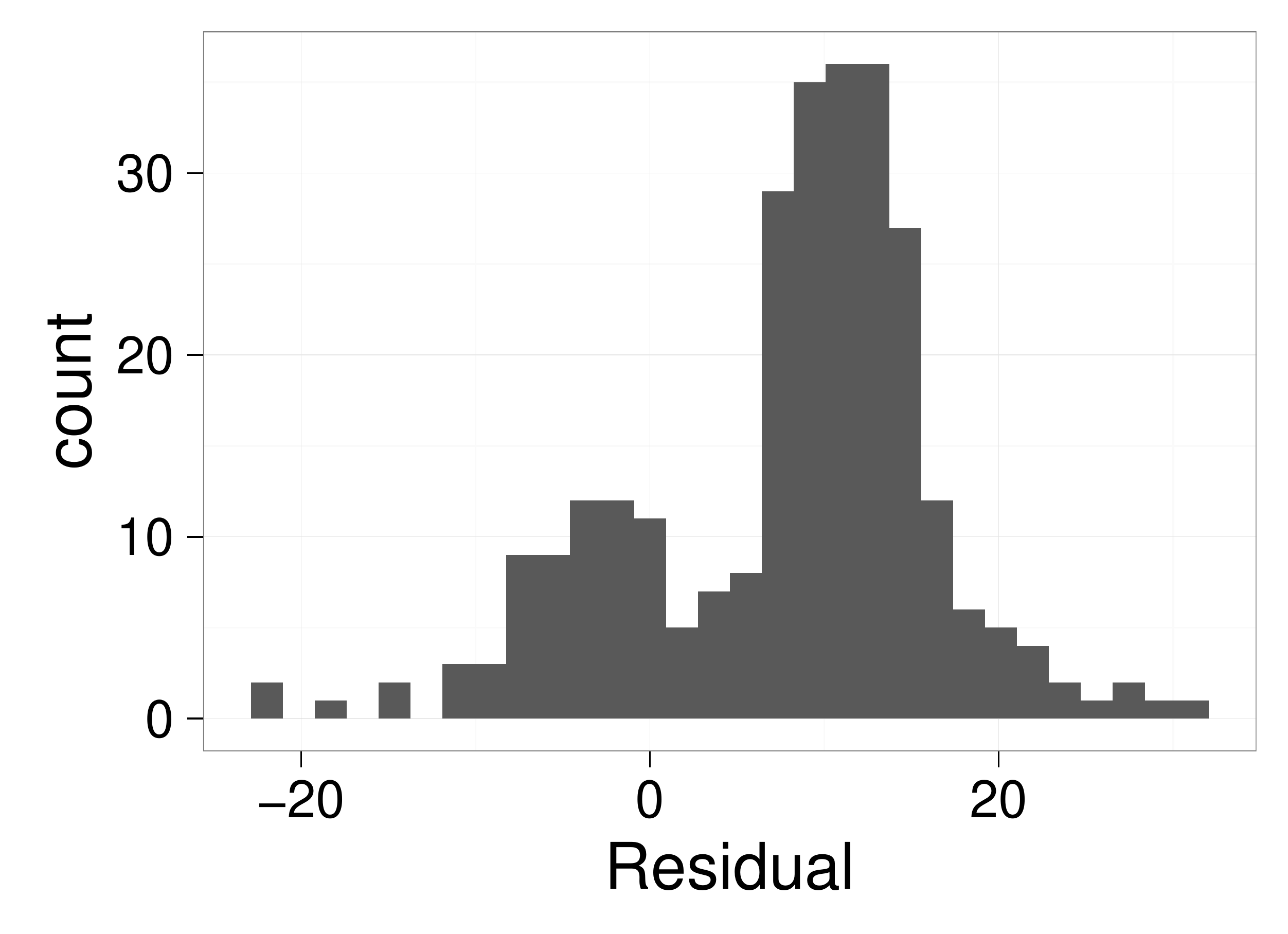}&
	\includegraphics[width=0.3\textwidth]{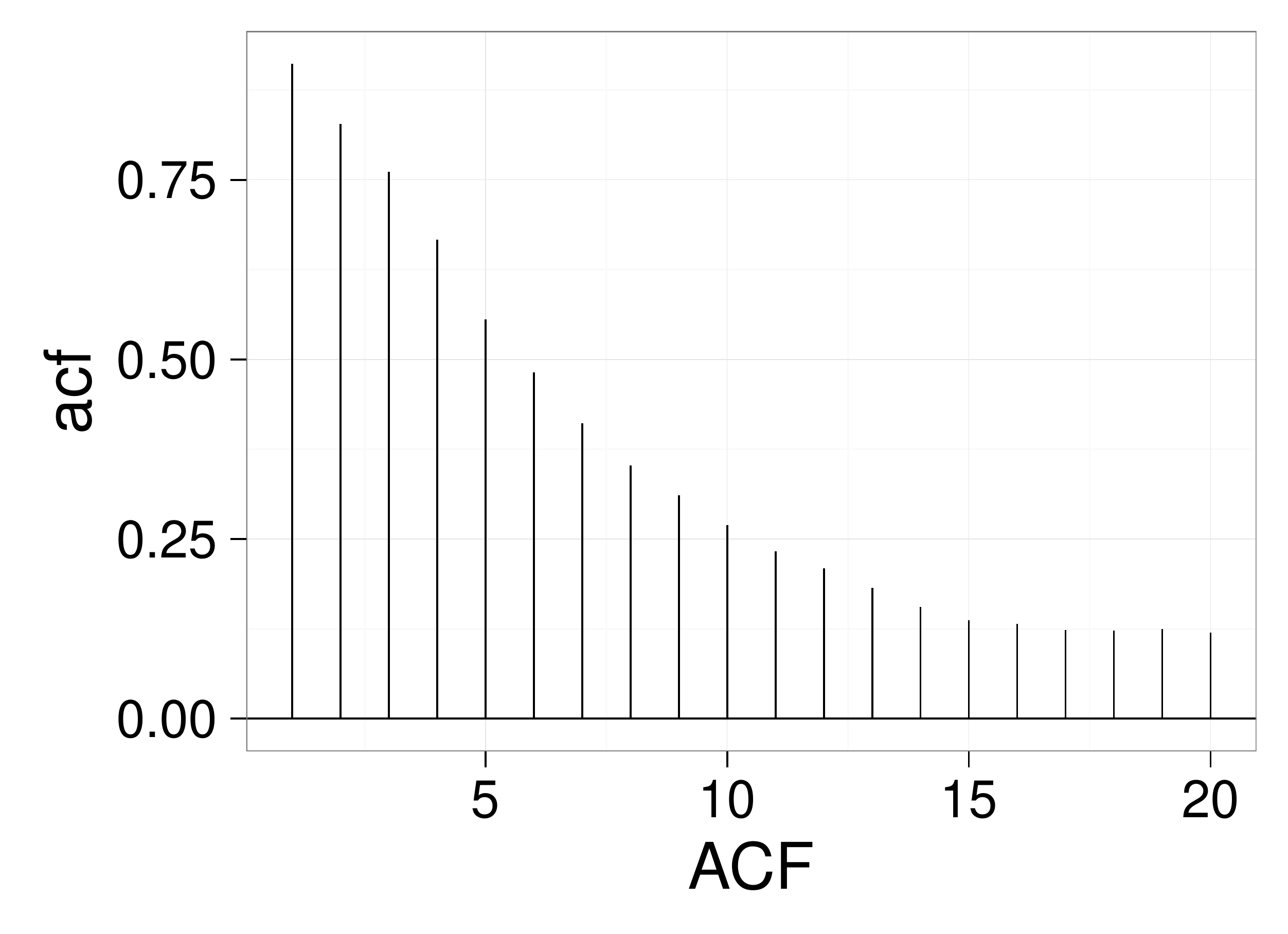}\\
	\rotatebox[origin=lt]{90}{DLM8}&
	\includegraphics[width=0.3\textwidth]{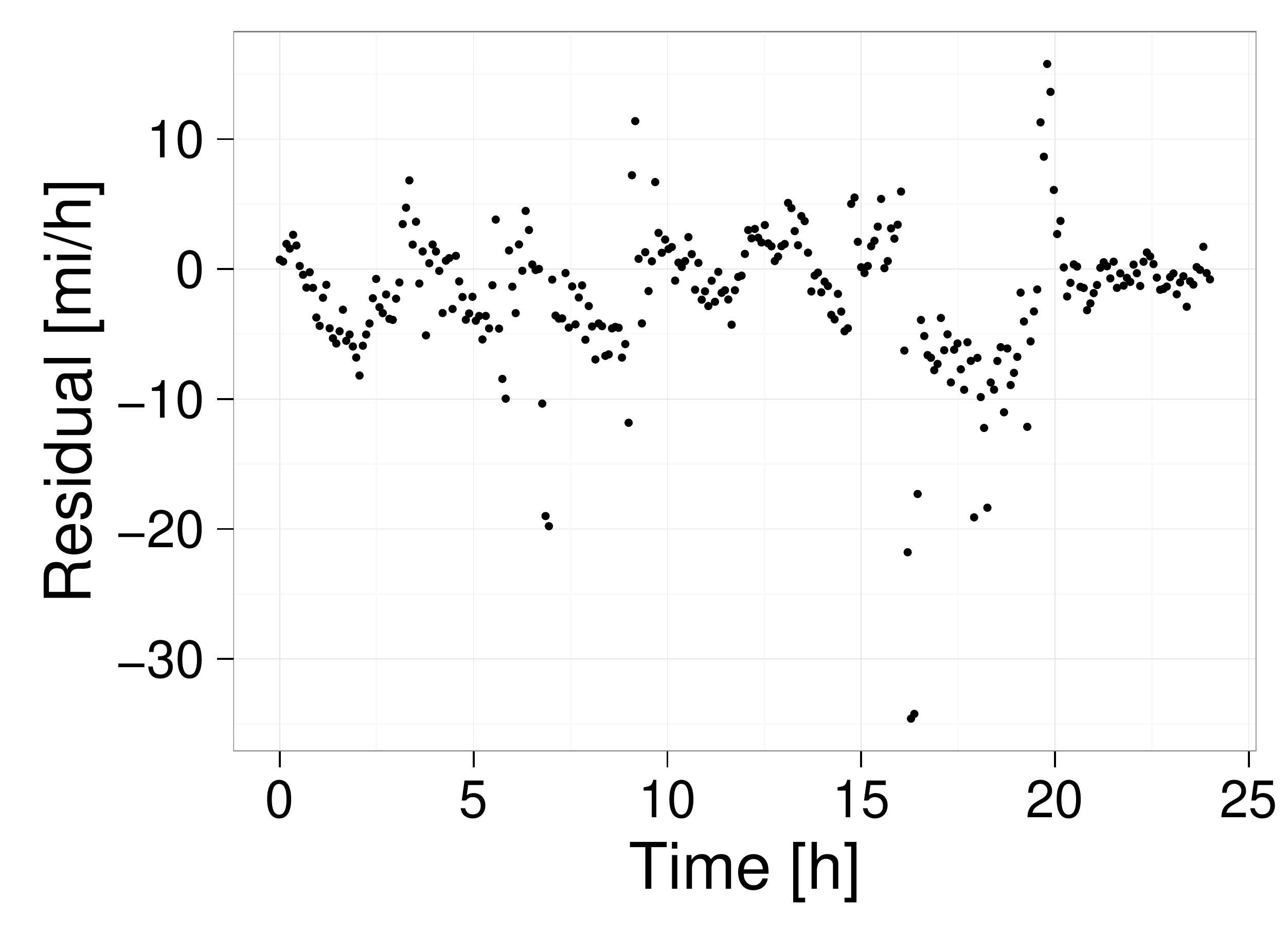}&
	\includegraphics[width=0.3\textwidth]{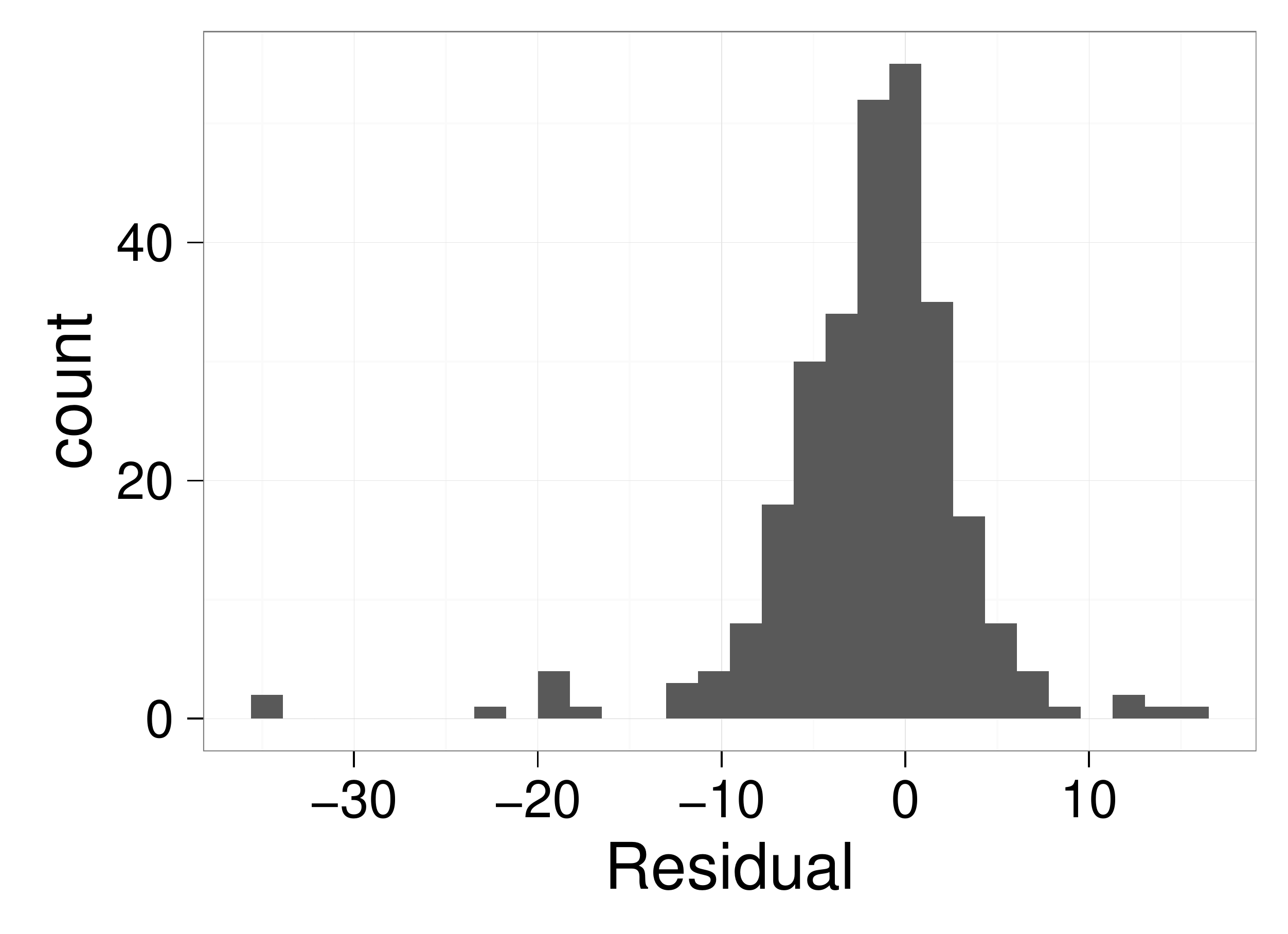}&
	\includegraphics[width=0.3\textwidth]{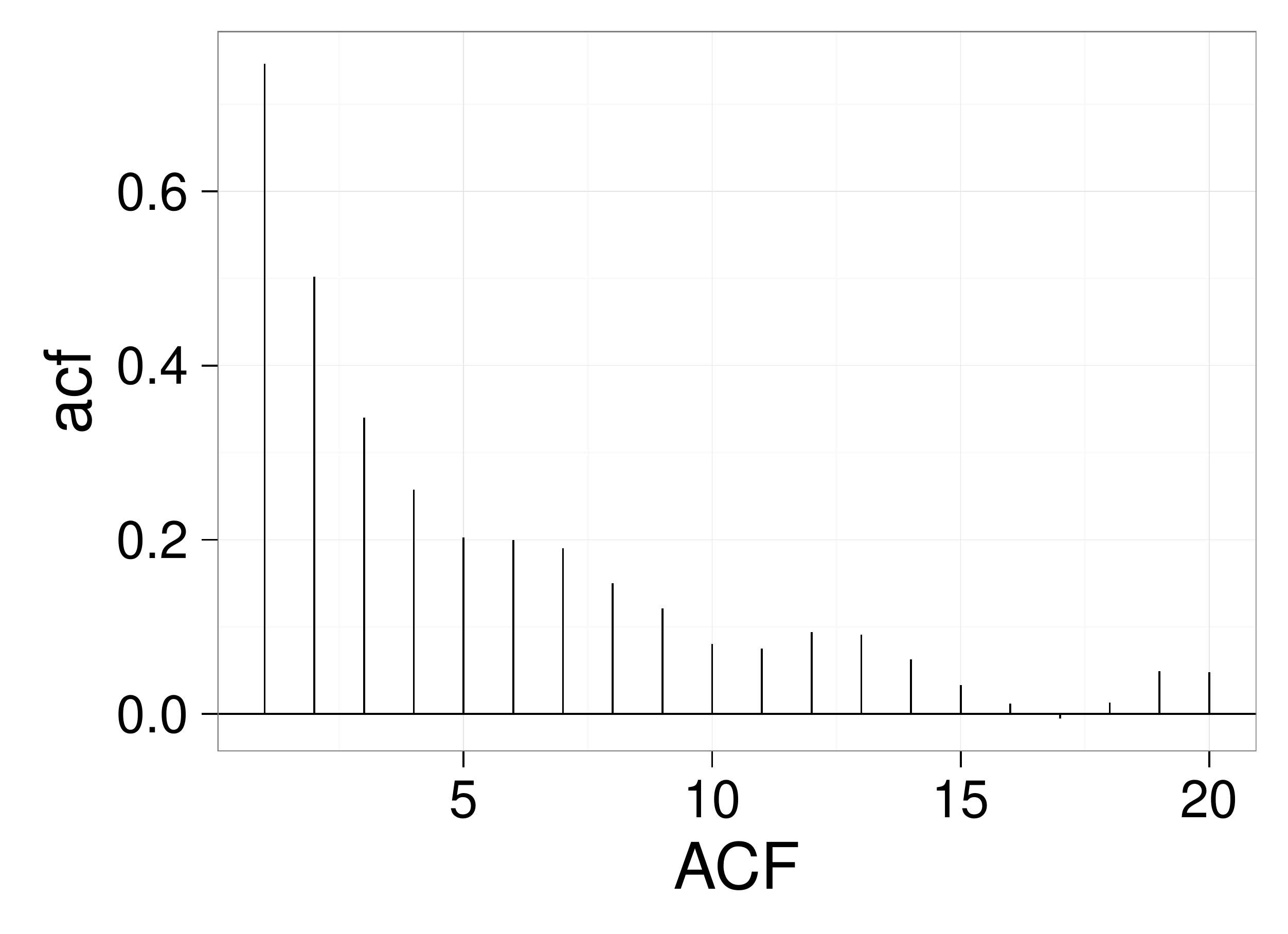}\\
		&(a) Time plot & (b) Histogram & (c) Autocorrelations\\ 	
\end{tabular}
\caption{Residual diagnostics plots for both vector auto-regressive and deep learning models. All plots are for residuals for data from July 29, 2013 (Monday).}
\label{fig:res}
\end{figure}

Figure~\ref{fig:res} shows that both DLM8 and VARM8 models do not account for all available information, they have autocorrelations in the residuals, and there is a structure in the data, that is not exploited by the models. From the autocorrelations plots  we can see that deep-learning model residuals are less correlated. Time plots show that deep learning residuals have less patterns and more uniform variance, compared to a VAR model. The histograms suggest that VAR residuals do not follow a normal distribution and DL do. Both of the models are biased, with mean residual for DL model being -2 and mean residual for VAR model being -7. 

Results of formal residual tests are shown in Table \ref{tab:res}.

\begin{table}[H]
	\centering
\begin{tabular}{l|l|l|l}
Test & NULL & VARM8 & DLM8\\\hline
Breusch-Godfrey & no autocorrelations & 4.27 (0.005)& 1.8 (0.15)\\
Box-Pierce & no autocorrelations & 995.77 (0)& 323.64 (0)\\
Breusch-Pagan & homoscedasticity & 6.62 (0.037)& 4.08 (0.13)\\
Lee-White-Granger & linearity in mean & 2.1 (0.13)& 0.16 (0.85)\\
Dickey-Fuller & non-stationary & -3.0154 (0.15)& -3.39 (0.05)\\
\end{tabular}
\caption{Results of the formal statistical tests for for autocorrelation, non-linearity,  homoscedasticity and stationarity in model residuals. The table shows value of the corresponding test statistic and p-value (in parentheses)  }
\label{tab:res}
\end{table}

A classical  Box-Ljung test~\cite{box1970distribution} for autocorrelation  shows that both models produce autocorrelated residuals. The $Q$-statistic, which is an aggregate measure of autocorrelation, is much higher for VAR model. However, as pointed out in previous neural networks applications to traffic analysis \cite{vlahogianni2013testing}, statistical tests based on Lagrange multiplier (LM) are more appropriate for the residual analysis. \cite{marcelo2006}, for example, notes that the asymptotic distribution of the test statistics in Box-Ljung test is unknown when a neural network model is used. Thus, we use a more appropriate LM-based Breusch-Godfrey test~\cite{hayashi2000econometrics} to detect the autocorrelations, Breusch-Pagan test for homoscedasticity and Lee-White-Granger~\cite{lee1993testing} test for  linearity in “mean” in the data. The LM-based tests suggest that the residuals from the deep learning model are less correlated, and more homoscedastic when compared to VAR model residuals. Though, we have to accept the linearity Null hypothesis for both models, according to  Lee-White-Granger, the $p$ value is much higher for the deep learning model. Another important finding is that the residuals are stationary for the DL model and are non-stationary for the VAR model. The formal Augmented Dickey-Fuller (ADF) test produced $p$-value of 0.06 for DL model and 0.15 for VAR model, with alternative hypothesis being that data is stationary. Stationary residuals mean that the model correctly captures all of the trends in the data.

Kolmogorov-Smirnov test shows that neither DL nor VAR model residuals are normally distributed. On the other hand, the residuals from DL model are less biased and are less correlated.


\subsection{Comparison to a single layer neural network }
Finally, deep learning is compared with a simple neural network model with one hidden layer. We used median filtering with a window size of 8 measurement (M8) as a preprocessing technique and predictors are pre-chosen using a sparse linear estimator. The in sample $R^2$ is 0.79 and MSE is 9.14. The out of sample metrics are 0.82 and 9.12.

%
%

%
%

\begin{figure}[H]
	\begin{tabular}{ccc}
		\includegraphics[width=0.33\textwidth]{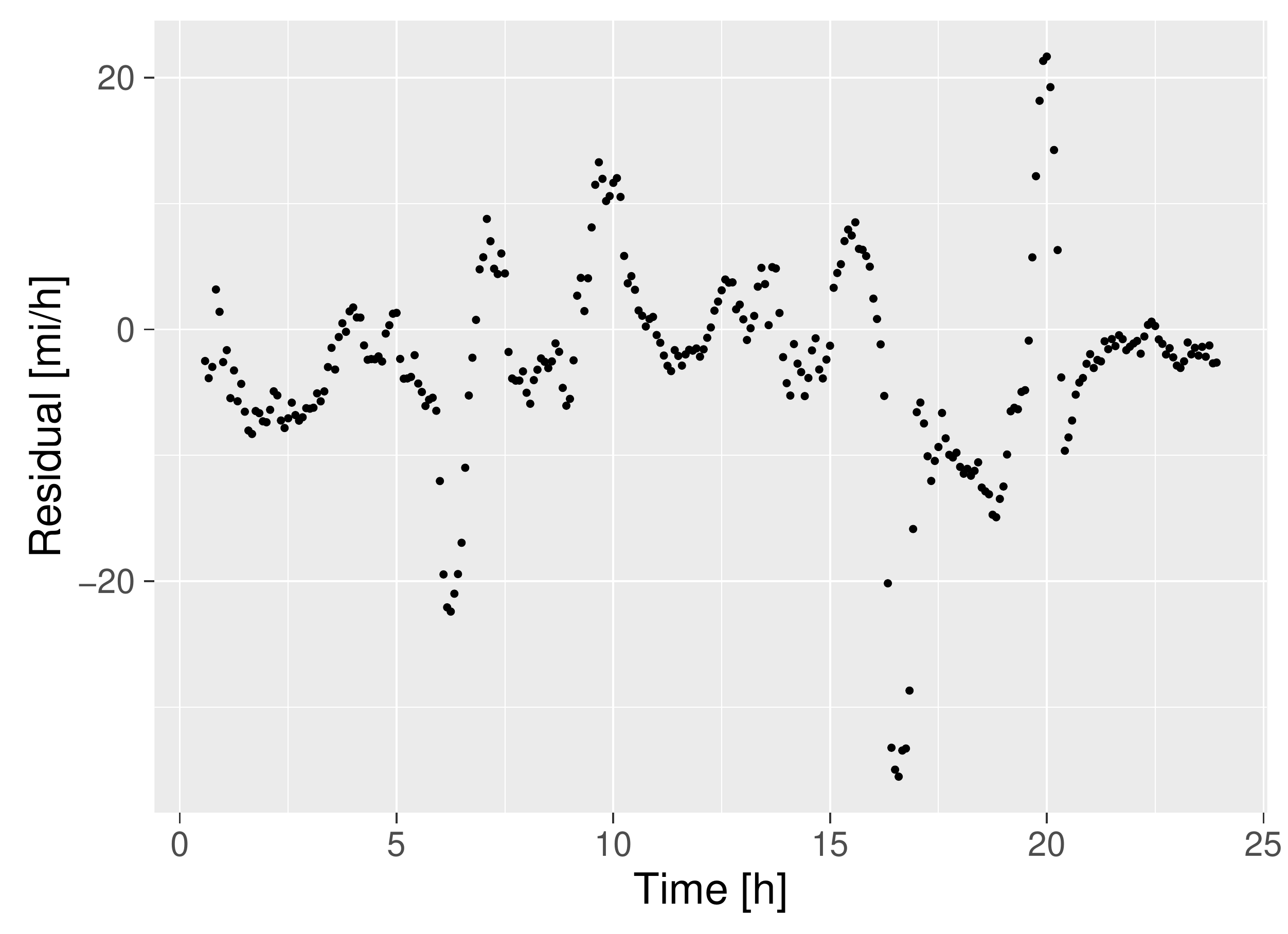}&
		\includegraphics[width=0.33\textwidth]{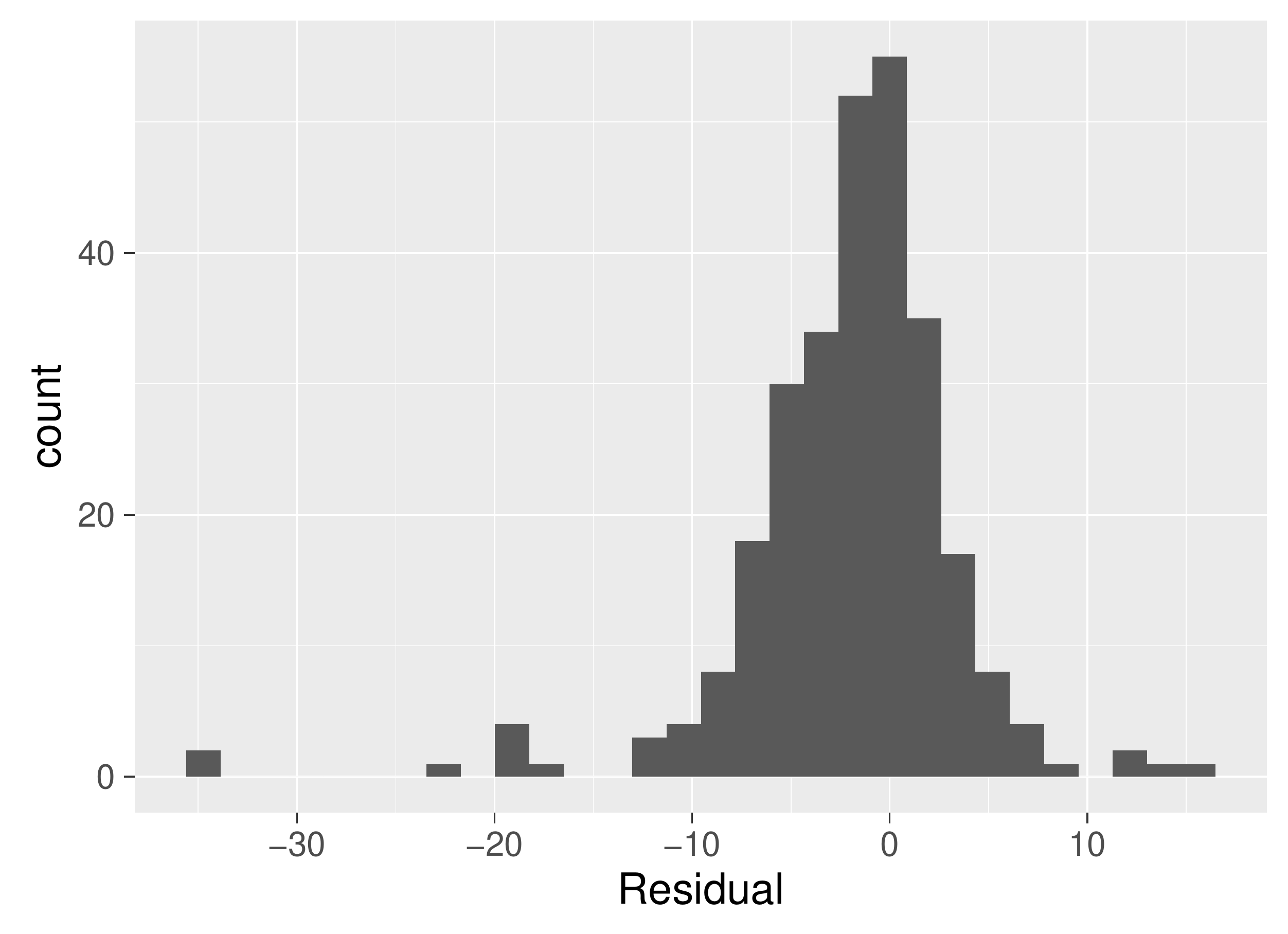}&
		\includegraphics[width=0.33\textwidth]{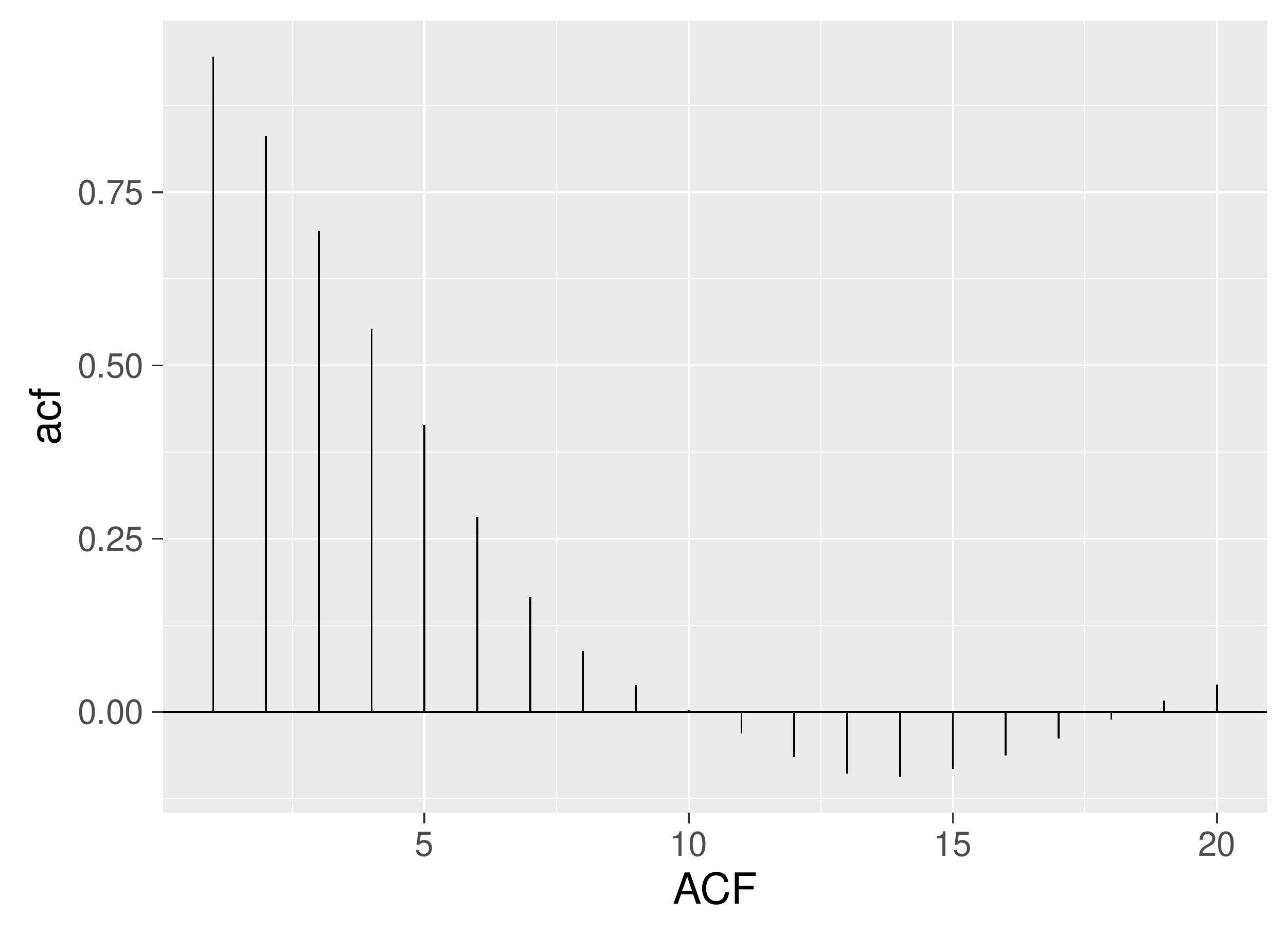}\\
		(a) Time plot & (b) Histogram & (c) Autocorrelations\\ 	
	\end{tabular}
	\caption{Residual diagnostics plots for neural network with one hidden layer. All plots are for residuals for data from July 29, 2013 (Monday).}
	\label{fig:res_1layer}
\end{figure}

The performance of the neural network model with one hidden layer is slightly worse than the one of the best linear model (VARM8L), the MSE is 
4\% higher and 
14\% higher when compared to the deep learning model (DLM8L). As shown in Figure~\ref{fig:res_1layer}, there is excess correlation structure left in the residuals, when compared to DLM8L. The model bias is comparable to the deep learning model and equals to  -2.1. A one-layer network model is less efficient and has less predictive power when compared to the deep learning network for traffic data.

\section{Discussion}\label{sec:discussion}

The main contribution of this paper is development of an innovative deep learning architecture to predict traffic flows. The architecture combines a linear model that is fitted using $\ell_1$ regularization and a sequence of $\tanh$ layers. The first layer identifies spatio-temporal relations among predictors and other layers model nonlinear relations. The improvements in our understanding of short-term traffic forecasts from deep learning are two-fold. First, we demonstrate that deep learning provides a significant improvement  over linear models. Second, there are also other types of networks that demonstrated superior performance for time series data. For example, the  recurrent neural network (RNN) is a class of  network where connections between units can form a directed cycle. This creates an internal state that allows to memorize previous data. 
Another class of networks that are capable of motorizing previous data are the long short term memory (LSTM) network, developed in \cite{hochreiter1997long}. It is an artificial neural network structure that addresses a problem of the vanishing gradient problem. In a sense, it allows for longer memory and it works even when there are long delays. It can also handle signals that have periodic components of different frequencies. Long short term memory and recurrent neural networks outperformed other methods in numerous applications, such as language learning \cite{gers2001lstm} and connected handwriting recognition \cite{graves2009offline}.

We empirically observed from data that recent observations of traffic conditions (i.e. within last 40 minutes) are stronger predictors rather than historical values, i.e. measurements from 24 hours ago. In other words, future traffic conditions are more similar to current ones as compared to those from previous days. Thus, it allowed us to develop a powerful model by using recent observations as model features.

One of the drawbacks of deep learning models is low explanatory power. In  a recent review of  short term forecasting techniques~\cite{vlahogianni2014short}, model interpretability is mentioned as one of the barriers in adapting more sophisticated machine learning models in practice. The idea of a deep learning model is to develop representations of the original predictor vector so that transformed data can be used for linear regression. There is a large volume of literature on studying representations for different domain specific models. Perhaps, the more advanced research on that topic was done for Natural Language Processing problems \cite{turian2010word, luong2013better}. One example of such representations are word embeddings, which are vectors associated with each word that contain latent features of the word and capture  its syntactic and semantic properties. For example, Miklov and Zweig~\cite{mikolov2013linguistic} show that if we calculate induced vector representation, ``King - Man + Woman''  on vector representation of the corresponding words, we get a vector very close to ``Queen.'' In the context of traffic predictions, relating the representations of input vectors to the fundamental properties of traffic flow is an interesting and challenging problem which needs to be studied further. 


\bibliography{ref}
\end{document}